\documentclass[superscriptaddress,pre,aps,floatfix,preprint]{revtex4}
\usepackage[utf8x]{inputenc}
\usepackage{graphicx
}
\usepackage[utf8x]{inputenc}
\usepackage{graphicx, 
}
\usepackage{epstopdf}
\usepackage{amsmath}
\usepackage{amsfonts}
\usepackage{color}

\begin{document}

\title{ Self-consistent theory for  systems with mesoscopic fluctuations
}
\author{ A. Ciach}
\affiliation{Institute of Physical Chemistry,
 Polish Academy of Sciences, 01-224 Warszawa, Poland}
 \author{ W. T. Gozdz}
\affiliation{Institute of Physical Chemistry,
 Polish Academy of Sciences, 01-224 Warszawa, Poland}
 \date{\today} 

\begin{abstract}
 We have developed a theory for inhomogeneous systems that allows for incorporation of effects of mesoscopic fluctuations.
 A hierarchy of equations relating the correlation and direct correlation functions  for the local excess $\phi({\bf r})$
 of the volume fraction of particles $\zeta$
 has been obtained, and an approximation leading to a closed set of equations for the two-point functions has been introduced
 for the disordered inhomogeneous phase.
 We have solved numerically the self-consistent equations 
 for  one (1D) and three (3D) dimensional  models with short-range attraction and long-rannge repulsion (SALR).  
 Predictions for all the qualitative properties of the 1D model agree with the exact  results, 
 but only semi-quantitative agreement is obtained
 in the simplest version of the theory.
 The effects of fluctuations in the two considered 3D models are 
 significantly different, 
 despite very similar properties of these models  in the mean-field approximation.
 In both cases we obtain the sequence of large - small - large 
 compressibility for increasing  $\zeta$.  The very small compressibility is accompanied by the 
 oscillatory decay of correlations with the correlation length  orders of magnitude larger than the size of particles.
 Only in one of the two considered models for decreasing temperature 
 the small compressibility becomes very small and
 the large compressibility becomes very large, and eventually van
 der Waals loops appear.
 Further studies are necessary to determine the nature of the strongly inhomogeneous phase present for intermediate volume fractions in 3D.
\end{abstract}
\maketitle
\section{introduction}
Ionic systems have been studied for decades 
and significant progress has been achieved, to a great deal thanks
to the important contribution by George Stell. His studies concerned in particular the  restricted primitive model (RPM).
 In the RPM equisized hard cores carry charges with equal magnitude and are immersed in structureless solvent.
Already in 1976 George Stell predicted phase separation into ion-poor and ion-rich phases
with the associated critical point\cite{stell:76:0}.
In 1992 Stell published very strong arguments that despite the long-range of the Coulomb potential 
the critical point in the RPM belongs to the Ising university class
\cite{stell:92:0}. The Ising universalty class was confirmed later by
field-theoretic methods~\cite{ciach:00:0,ciach:05:0,ciach:06:1} and by 
the hierarchical reference and collective variables theories \cite{parola:11:0,patsahan:03:0}.  The Stell prediction was verified
by simulations  
 in 2002~\cite{luijten:02:0}. 

In reality the Coulomb potential often competes with the specific van der Waals interactions and, in particular in the case of 
charged globular proteins or colloid particles in complex solvents, with various solvent-induced effective interactions.
Moreover, the size, shape  and charge of the positively and negatively charged ions or particles can be  different. 
This difference is moderate in room temperature ionic liquids (RTIL), but in the case of charged globular proteins in solvents containing
 microscopic counterions
the size- and charge ratio becomes $\sim 10$, and it increases even to $\sim 10^2-10^4$ in the case of colloid particles.
The size- and charge asymmetry, as well as the competing non-Coulombic interactions both may lead to spacial inhomogeneities
on the length scale larger than the size of ions or charged particles. One of the first theoretial observations of the instability
of the homogeneous phase with respect to periodic charge- or number-density distribution
was made by Stell and coworkers~\cite{ciach:01:1,ciach:06:0,ciach:07:0,patsahan:12:0}. Inhomogeneitis, 
in particular clusters, networks
or layers, or exotic crystals, were observed also in experiments and simulations
~\cite{arora:88:0,russina:11:0,desfougeres:10:0,elmasri:12:0,royall:04:0,weis:98:0,stradner:04:0,campbell:05:0,blaaderen:05:0,hynninen:06:0,cheong:03:0,candia:06:0,toledano:09:0,kowalczyk:11:0}. 
Instability with respect to periodic distribution of particles in space competes with the phase separation, and may lead to quite complex phase 
behavior~\cite{ciach:03:0,ciach:05:0,ciach:06:0,ciach:07:0,patsahan:12:0}.

 Despite the progress in studies of ionic systems and charged particles,
many important questions remain open, because
accurate description of systems with mesoscopic inhomogeneities remains a challenge. 
Liquid matter theories, such as the generalized mean-spherical approximation
or the self-consistent Ornstein-Zernike approximation (SCOZA)~\cite{dickman:96:0}
work very well for simple liquids. In particular, SCOZA yields globally accurate phase diagram~\cite{pini:98:0}.
Theories like SCOZA and hierarchical reference theory  (HRT) include fluctuations over all length scales;
 the limitation of SCOZA and HRT is that no solutions are
found when a uniform state becomes unstable with respect to a modulated phase~\cite{archer:07:0,pini:00:0}.
It is not easy to obtain solutions in the liquid-matter theories for inhomogeneous systems with competing interactions,
such as the short-range attraction 
and long-range repulsion potential (SALR).
Nevertheless, some features due to the SALR potential
can and have been studied~\cite{archer:07:0,pini:08:0,pini:00:0}. In particular, enhanced density fluctuations, a tendency towards cluster
formation and a growth of the compressibility 
in a large density and temperature interval
close to the liquid-vapour transition
was  observed by Pini et al. in the case of weak repulsion~\cite{pini:00:0,pini:08:0}.


The inhomogeneities  that occur at the microscopic length scale near external surfaces are successfuly 
described by the density functional theory (DFT)~\cite{evans:79:0}. However,
in the case of spontaneously occurying mesoscopic inhomogeneities the predictions of the DFT deviate
from the results of simulations more significantly than in simple fluids~\cite{archer:08:0,imperio:06:0}. This is 
because fluctuations, such as displacements, reshaping, merging or splitting of the aggregates play an important role in these systems.

The fluctuations can be taken into account in the Brazovskii field theory (BFT)~\cite{brazovskii:75:0} relatively easily,
but since this theory is of phenomenological nature, neither the equation of state (EOS) nor the phase diagram in terms of real
theromdynamic variables can be determined. 
 Note that the  displacements, reshaping, merging or splitting of the aggregates
 can appear either spontaneously or as a result of external stimuli,
 and one can expect that thermodynamic susceptibilities, in particular the compressibility and the specific heat, are much different 
 than in homogeneous systems. Since the complexity of these systems leads to serious technical difficulties,
 their structural, mechanical and thermal properties are not yet fully understood.
 
 The aggregates are periodically distributed in space in the ordered phases analogous to lyotropic liquid crystals, 
 but in the {\it inhomogeneous
 disordered phase}, analogous to microemulsion, the aggregates are ordered only locally. The SALR and amphiphilic systems
  have similar properties, 
 since as shown in Ref.\cite{ciach:13:0}, both can be described by the Brazovskii functional. (We should mention here that
  George Stell studied the amphiphilic systems as well, in particular in Ref.~\cite{ciach:88:0}.)
 However, while the lyotropic liquid crystals in amphiphilic systems are quite common~\cite{latypova:13:0},
 only disordered distribution of spherical or elonogated clusters  was observed
 in the SALR system by confocal microscopy~\cite{stradner:04:0,campbell:05:0}. It is unclear if the observations concerned the 
 disordered inhomogeneous phase, or just one state of the ordered phase  
 and after averaging the distribution of the aggragates would be periodic.  The ordered hexagonal and lamellar phases were obtained by
 molecular dynamics simulations, but the spherical clusters were not ordered periodically in this simulation~\cite{candia:06:0}.
{\bf In recent MC simulations \cite{zhuang:16:0} a similar sequence of phases as in Ref.\cite{ciach:13:0} was obtained.}
 
\begin{figure} 
\includegraphics[scale=0.34]{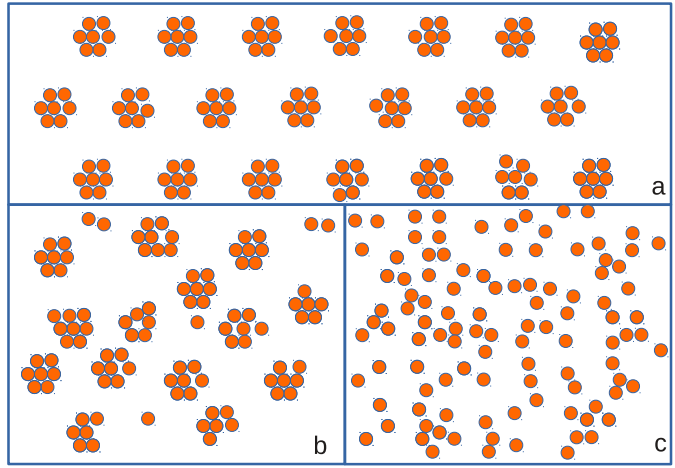} 
\begin{center}
 \caption{ Panel (a) shows a typical configuration in the {\it ordered inhomogeneous} phase, where clusters of particles
 form a hexagonal lattice. In panel (b) a typical configuration in the  {\it disordered inhomogeneous} phase is shown. 
 The particles self-assemble into clusters, 
 but the clusters are not distributed periodically. Fluctuations of the positions lead to the average volume fraction 
 that is position-independent. Properties of such an inhomogeneous system can differ from the 
  disordered phase that is homogeneous
 on the mesoscopic length scale (panel (c)) despite the same value of the average volume fraction of the particles.
In this work we are concerned with properties of systems shown in panel (b), where mesoscopic fluctuations play a dominant role.
 }
 \label{f0}
 \end{center}
 \end{figure}

 In Fig.\ref{f0} we show a cartoon with schematic representation of
 a typical distribution of particles in the periodically ordered inhomogeneous phase, and in the disordered phase
 that is either inhomogeneous (Fig.\ref{f0}b) or homogeneous (Fig.\ref{f0}c) on the mesoscopic length scale. 
 “The disordered phase” and “the homogeneous phase” are often treated as synonyms for the phase with the position-independent average density.
 However, the position-independent average density does not necessarily mean that the structure is homogeneous at the mesoscopic length scale 
 in the majority of states (see Fig.\ref{f0}b).
 Here we call the phase with position-independent average density ``the disordered phase''. 
 
 The ordered periodic phases in the SALR and amphiphilic systems are quite well described by the mean-field (MF)
 theories~\cite{archer:08:0,ciach:10:1,gozdz:96:1}. In the disordered inhomogeneous phase, however, the mesoscopic fluctuations play
 a key role, and its properties are not correctly predicted on the  MF level.  
 The aggragates are clearly seen in confocal microscope and in simulation snapshots, but
 due to fluctuations, the average volume fraction of the
 particles, $\bar\zeta({\bf r})$, is position-independent, $\bar\zeta({\bf r})=const.$.
 For this reason one cannot distinguish the inhomogeneous and homogeneous structure of the disordered phase based solely
 on the one-particle distribution function.
 In this work we propose a theoretical method of investigation
 of the disordered inhomogeneous phase, and apply the formalism to  a one-dimensional (1D) and a three-dimensional (3D) 
 SALR model with strong repulsion in order to calculate the EOS.
 
 Some information about properties of the disordered inhomogeneous phase was obtained 
 in Ref.\cite{pekalski:13:0}, where a 1D
 lattice model with nearest-neighbor 
 attraction and third-neighbor repulsion was solved exactly.
 When the repulsion is strong
 enough, then clusters consisting of three particles separated by three empty sites 
 (i.e.  ...ooo...ooo...ooo...ooo... where ``.'' and ``o''
 denote the empty and ocuppied site repectively) are energetically favourable. 
 The volume fraction of particles in this ordered 
 structure, stable only at  temperature $T=0$, is $\zeta=1/2$. 
 There are no phase transitions in 1D models with short-range interaction, but 
 at low $T$ and $\zeta\approx 1/2$ the disordered phase turns out to be strongly inhomogeneous. 
 In Ref.\cite{pekalski:13:0} it was found that for $T>0$ 
 the correlation function exhibits an oscillatory decay with the period $\approx 6$. 
 For $\zeta\approx 1/2$ the correlation length increases to very large values and the compressibility decreases to very small values 
 for decreasing $T$. The compressibility is small for  $\zeta\approx 1/2$, because the 
 increase of $\zeta$ leads to shorter separation between the clusters
 and to repulsion between them. Large compressibility was obtained for $\zeta$ that is either too small or too large for formation of the 
 periodic structure that is favourable energetically. 
 The  chemical potential $\mu(\zeta)$ and pressure 
 $p(\zeta)$ isotherms at low $T$ are much different than in simple fluids. When $T$ decreases, the slope of $\mu(\zeta)$ 
 becomes very small 
 for a range of both small and large 
 $\zeta$, and very large  for $\zeta\approx 1/2$. The very small slope of $\mu(\zeta)$ was interpreted as pseudo phase transition 
 between homgeneous very dilute or very dense phase, and the inhomogeneous phase present for $\zeta\sim 1/2$. 
 Interestingly, for $\zeta\sim 0.6$ the pressure 
 decreases for increasing $T$, in contrast to simple fluids. 
 Properties of the two-dimensional and  three-dimensional (3D) systems can be easily determined neither 
 by exact calculations nor by simulations, due to large finite size effects and the  collective motion of the aggregates. 
 For this reason
 it is important to develop an approximate predictive theory and   test its accuracy by comparison with 
 the exact results obtained in the 1D model.
 
 Development of a tractable theory that could allow for quantitative or at least semi-quantitative description of
 structural, mechanical and thermal properties of systems with mesoscopic inhomogeneities is our long-term goal.
 In Ref.\cite{ciach:08:1,ciach:11:0,ciach:13:0, ciach:15:0} we have made first steps in this direction.
 The general formalism allows for investigation of disordered and periodically ordered phases, but the obtained equations
 are very difficult and in practice approximations are necessary. In order to verify  quality of various approximations,
 one should compare the results with exact solutions that so far exist only for the disordered phase~\cite{pekalski:13:0}. 
 Before considering the modulated phases, we limit ourselves to the disordered phase to verify various approximate schemes.
 
 In this work we further develop
 our theory that combines liquid-matter theory, DFT and BFT methods. The present version of the approximate theory 
 is based to a large extent on the derivation described in Ref.\cite{ciach:15:0}, and summarized briefly in sec.2. 
 We focus on  a one-component system. 
 In the context of charged particles this means that we integrate out the degrees of freedom of
 the counterions, and consider screened electrostatic potential between the charged particles. The explicit counterions
 can be considered within our formalism at a later stage.
 
The self-consistent Gaussian appoximation developed in Ref.\cite{ciach:15:0} is rather simple and yields results that agree 
qualitatively with majority of the exact results obtained for the 1D lattice model~\cite{pekalski:13:0}. 
Unfortunately, it turned out that in the case of the 3D
SALR model in continuum space our equation for the direct correlation function has no solutions for the phase space region where
inhomogeneities are expected. It resembles the above  mentioned lack of solution in the SCOZA.
To overcome this problem, in this work we develop a theory beyond the Gaussian approximation.
In sec.3 a hierarchy of equations relating the many-point correlation and direct correlation functions for mesoscopic volume fraction
is constructed.
We make an approximation for the direct four-point correlations, and obtain an
equation relating the two-point correlation and direct correlation 
functions that  together with the Ornstein-Zernike (OZ) equation form a closed set of equations. Finally, 
we obtain expressions for $\mu(\zeta)$ and $p(\zeta)$ which contain contributions resulting from the mesoscopic fluctuations. 
In sec.4 the results obtained in the approximate theory for the 1D lattice model are compared
with the exact results of Ref.~\cite{pekalski:13:0}. The agreement is
 much better than in the Gaussian approximation. In sec.5 we present results of our theory for the 3D SALR model with both the 
attractive and the repulsive part of the interaction potential having the Yukawa form. We choose two sets of parameters, both favouring
periodic distribution of particles over the homogeneous state.
In the first model the attraction range 
is very short and the repulsion barrier is small. For this potential, small clusters are formed and
the gas-liquid separation is energetically unfavourable compared to the homogeneous state.
In the second model the attraction range and the repulsion 
barrier are both larger. Larger clusters are formed and the gas-liquid separation is energetically 
favourable over the homogeneous state.
In MF approximation the phase diagrams of the two models are very similar~\cite{ciach:10:1}.
Here, we ask if the effects of fluctuations 
on the shape of the $\mu(\zeta)$ and $p(\zeta)$ curves depend on the range and amplitude of
the attractive and the repulsive part of the 
interaction potential. Sec.6 contains summary and discussion.

 \section{Brief summary of the theory for systems with mesoscopic inhomogeneities}
 We consider systems with inhomogeneities  on a length scale
significantly larger than the size of  molecules $\sigma\equiv 1$ (see Fig.\ref{f0}b).
In our theory~\cite{ciach:08:1,ciach:11:0,ciach:15:0} a
mesoscopic volume fraction is described by a smooth function
$\zeta({\bf r})$ equal to
the fraction of the volume of the mesoscopic region with a center at ${\bf r}$ 
that is covered by the particles. By fixing the mesoscopic state given by
$\zeta({\bf r})$ we impose a constraint on the available microstates. The grand thermodynamic potential 
in the presence of this constraint
is  denoted by $\Omega_{co}[\zeta({\bf r})]$. After removal of the constraint  mesoscopic fluctuations may appear, and 
the grand potential is given
by~\cite{ciach:08:1,ciach:15:0}
\begin{equation}
\label{Om}
 \Omega=\Omega_{co}[\bar\zeta]-k_BT\ln\Bigg(\int D\phi e^{-\beta H_f}\Bigg)
\end{equation}
with $\beta^{-1}=k_BT$,  $k_B$ the Boltzmann constant, and
\begin{equation}
\label{Hf}
 H_f[\bar\zeta,\phi]=\Omega_{co}[\bar\zeta+\phi]-\Omega_{co}[\bar\zeta],
\end{equation}
where $\bar\zeta$ denotes the average volume fraction, and
\begin{equation}
 \phi:=\zeta-\bar\zeta
\end{equation}
is the mesoscopic fluctuation. The first term in (\ref{Om}) contains contributions from the 
fluctuations on the microscopic length scale in the absence of mesoscopic fluctuations. 
The second term contains the contributions from the fluctuations on the mesoscopic 
length scale, i.e. from different mesoscopic inhomogeneities that are thermally excited with the probability $ e^{-\beta H_f}/\Xi$.
When $\bar\zeta$ is the average volume fraction, then it follows that $\langle\phi\rangle=0$, where
\begin{equation}
\label{Xav}
 \langle X \rangle:=\Xi^{-1}\int D \phi X e^{-\beta H_f}
\end{equation}
and
\begin{equation}
\label{Xi}
 \Xi=\int D \phi  e^{-\beta H_f}.
\end{equation}
In the following  $\langle\phi\rangle=0$ is always assumed. 

We introduce the functional derivatives
\begin{eqnarray}
\label{Cn}
 C_n({\bf r}_1,...,{\bf r}_n):=\frac{\delta^n \beta\Omega}{\delta\bar\zeta({\bf r}_1)...\delta\bar\zeta({\bf r}_n)}=
 \\
 \nonumber
 C_n^{(0)}({\bf r}_1,...,{\bf r}_n)
 -\frac{\delta^n}{\delta\bar\zeta({\bf r}_1)...\delta\bar\zeta({\bf r}_n)}\ln\Big(\int D\phi e^{-\beta H_f[\bar\zeta,\phi]}
 \Big)
\end{eqnarray}
where
\begin{equation}
 C_n^{(0)}({\bf r}_1,...,{\bf r}_n)=\frac{\delta^n \beta\Omega_{co}}{\delta\bar\zeta({\bf r}_1)...\delta\bar\zeta({\bf r}_n)}.
\end{equation}
  $C_n$ and $C_n^{(0)}$ are functionals of $\bar\zeta$.
 
  For $n=1,2$ we obtain from (\ref{Cn})
 \begin{equation}
 \label{C1}
C_1({\bf r})=C_1^{(0)}({\bf r}) +
 \langle \frac{\delta \beta H_f}{\delta \bar\zeta({\bf r})}\rangle,
\end{equation}
and
\begin{equation}
\label{C2}
 C_2({\bf r}_1,{\bf r}_2)=C_2^{(0)}({\bf r}_1,{\bf r}_2) +
 \langle \frac{\delta^2 \beta H_f}{\delta \bar\zeta({\bf r}_1)\delta \bar\zeta({\bf r}_2)}\rangle 
 - \langle\frac{\delta \beta H_f}{\delta \bar\zeta({\bf r}_1)}\frac{\delta \beta H_f}{\delta \bar\zeta({\bf r}_2)}\rangle^{con}
\end{equation}
where  
\begin{equation}
 \langle X({\bf r}_1) Y({\bf r}_2) \rangle^{con}:
 =\langle X({\bf r}_1) Y({\bf r}_2) \rangle -\langle X ({\bf r}_1) \rangle\langle Y({\bf r}_2)  \rangle.
\end{equation}
 The  explicit expressions for $C_3$ and $C_4$ are given in Appendix A.

 We make
the standard  local MF approximation for the grand potential with suppressed mesoscopic fluctuations,

\begin{equation}
\label{Omco}
 \Omega_{co}[\zeta]=U[\zeta] -TS[\zeta]-\mu N[\zeta],
\end{equation}
where  $U[\zeta],S[\zeta]$  and  $N[\zeta]$ are the internal energy, the entropy and
 the number of particles respectively
 in the system with the mesoscopic volume fraction constrained to have the form $\zeta$. 
The entropy and the internal energy  in the local density approximation are given by 
\begin{equation}
-TS[\zeta]=\int d{\bf r} f_h(\zeta({\bf r}))
\end{equation}
and
\begin{equation}
\label{U}
 U[\zeta]=\frac{1}{2}\int d{\bf r}_1\int d{\bf r}_2\zeta({\bf r}_1)V({\bf r}_1-{\bf r}_2)\zeta({\bf r}_2).
\end{equation}
In order not to include the contributions to the internal energy from overlapping hard cores of the particles, we assume 
 $V({\bf r}_1-{\bf r}_2)=u({\bf r}_1-{\bf r}_2)\theta(|{\bf r}_1-{\bf r}_2|-1)$, with $u$ denoting
the interaction potential.
We use volume fraction rather than density in (\ref{U}), therefore  we should re-scale 
the interaction potential $u({\bf r}_1-{\bf r}_2)$ by the factor $(6/\pi)^2$ to obtain the same energy
as in the standard theory. 
We also re-scale the chemical potential, $\bar\mu=(6/\pi)\mu$, so that 
\begin{equation}
\label{muN}
 \mu N[\zeta]=\bar\mu\int d{\bf r}\zeta({\bf r}).
\end{equation}
%
For $\Omega_{co}$ defined in (\ref{Omco})-(\ref{muN}) we have 
\begin{equation}
\label{C01}
  C_1^{(0)}({\bf r})= \int d{\bf r}_1 \zeta({\bf r}_1)\beta V({\bf r}_1-{\bf r})+A_1(\zeta({\bf r}))-\beta \bar\mu
\end{equation}
\begin{equation}
\label{C02}
 C_2^{(0)}({\bf r}_1,{\bf r}_2)=\beta V({\bf r}_1-{\bf r}_2)+A_2(\zeta({\bf r}_1))\delta({\bf r}_1-{\bf r}_2)
\end{equation}
and for $n\ge 3$
\begin{equation}
\label{C0n}
 C_n^{(0)}({\bf r}_1,...,{\bf r}_n)=A_n(\zeta({\bf r}_1))\delta({\bf r}_1-{\bf r}_2)...\delta({\bf r}_{n-1}-{\bf r}_n),
\end{equation}
where
\begin{equation}
\label{An}
 A_n(\zeta)=\frac{d^n\beta f_h(\zeta)}{d \zeta^n}.
\end{equation}

With the above form of $\Omega_{co}$ 
we can obtain $\bar\zeta$ for given $T$ and $\bar\mu$ from  the requirement that $\Omega$ takes the minimum, i.e. 
$C_1({\bf r})=0$, with  $C_1$ given in Eq.(\ref{C1}).
The only difficulty is  the calculation of the fluctuation correction to the mean-field (MF) equation $C_1^{(0)}=0$. 
In order  to calculate this correction,
 it is neccessary to perform functional integrals (see  (\ref{Xav}) and (\ref{C1})).
In practice only   the Gaussian functional integrals can be calculated.
In order to perform the functional integrals in (\ref{Om}), (\ref{C1}) and (\ref{C2}), 
we have to approximate $H_f$ by an effective functional that is quadratic in the fluctuation $\phi$, or
expand $H_f[\bar\zeta,\phi]$ in a functional Taylor series w.r.t. $\phi$.

In Ref.\cite{ciach:15:0} we have made the approximation 
\begin{eqnarray}
\label{HG}
 \beta H_f\approx\beta  H_G= \frac{1}{2}\int d{\bf r}_1\int d{\bf r}_2 \phi({\bf r}_1)C_2({\bf r}_1,{\bf r}_2)\phi({\bf r}_2)
\end{eqnarray}
and  obtained from
(\ref{C2}) the following approximate equation:
%
%
\begin{equation}
\label{selfcon}
 C({\bf r}_1,{\bf r}_2)\approx C^{(0)}_2({\bf r}_1,{\bf r}_2)+\frac{A_4(\bar\zeta({\bf r}_1))}{2}G({\bf r}_1,{\bf r}_2)\delta({\bf r}_1-{\bf r}_2)
 -\frac{A_3(\bar\zeta({\bf r}_1))A_3(\bar\zeta({\bf r}_2))}{2}G^2({\bf r}_1,{\bf r}_2).
\end{equation}
where $G({\bf r}_1,{\bf r}_2):=\langle \phi({\bf r}_1) \phi({\bf r}_2)\rangle$ 
satisfies the  OZ equation
\begin{equation}
\label{CG}
 \int d {\bf r}_2 G({\bf r}_1,{\bf r}_2)C({\bf r}_2,{\bf r}_3)=\delta({\bf r}_1,{\bf r}_3) 
\end{equation}
and 
$C\equiv C_2$. From now on we omit the subscript $2$ in the case of the two-point correlation and direct
correlation functions to simplify the notation. 
Eqs.(\ref{selfcon}) and (\ref{CG}) have to be solved self-consistently. Eq.(\ref{selfcon}) 
is valid for any $\bar\zeta({\bf r})$, including the modulated phases.

The above Gaussian approximation is relatively simple, and correctly predicts  the main qualitative features of the one-dimensional 
lattice model that was solved exactly in Ref.~\cite{pekalski:13:0}. Unfortunately, it has a serious disadvantage.
In the case of a three-dimensional
SALR model Eq.(\ref{selfcon}) has solutions only for a limited range of  $T$ and $\bar\zeta$
in the disordered phase. 
This lack of solution occurs 
because  $A_3(\zeta)$ is large for small or large values of $\zeta$, and the RHS of Eq.(\ref{selfcon}) becomes negative. 
On the other hand, in Fourier representation $\tilde C(k)>0$ for high $T$ and never vanishes, since the RHS of Eq.(\ref{selfcon})
diverges for $\tilde C(k)=0$.
The lack of solution indicates that the self-consistent Gaussian approximation is oversimplified.
 Thus, if we want to study a three dimensional system, we have 
to abandon the elegant and simple Gaussian approximation.
\section{Beyond the Gaussian approximation}

In the first step beyond the Gaussian approximation we expand $H_f[\bar\zeta,\phi]$ 
in a functional Taylor series w.r.t. $\phi$,
\begin{equation}
\label{Taylor2}
 \beta H_f[\bar\zeta,\phi]=  \beta H_0+\beta \Delta H
\end{equation}
with 
\begin{equation}
\label{Taylor20}
 \beta H_0[\bar\zeta,\phi]=  \frac{1}{2}
  \int d{\bf r}_1 \int d{\bf r}_2 \phi({\bf r}_1)C_2^{(0)}({\bf r}_1,{\bf r}_2)\phi({\bf r}_2)
\end{equation}
and
\begin{equation}
\label{Taylor2int}
 \beta \Delta H[\bar\zeta,\phi]=  \int d{\bf r} C_1^{(0)}({\bf r})\phi({\bf r})
 +\sum_{n\ge 3}\int d{\bf r} \frac{A_n(\zeta({\bf r}))}{n!} \phi({\bf r})^n.
\end{equation}
In practice the Taylor expansion is truncated, and  only terms up to  $\phi^n$ are kept.
For stability reasons $n$ must be an even number. The truncation is justified 
provided that the neglected terms are smaller than the kept terms.
Since $A_{2m+1}$ vanishes for some value of $\zeta$,  we compare the terms $\propto \phi^{n}$ and $\propto \phi^{n+2}$, and 
require that $A_n\phi^n/n!>A_{n+2}\phi^{n+2}/(n+2)!$ for even $n$.
 The above necessary condition 
 must be satisfied by the fluctuations $\phi$ that yield the major
contribution 
to the average quantities. If for such fluctuations the above condition is satisfied, then it should be satisfied
by $\langle\phi({\bf r})^2\rangle$.

In this work we develop a lowest-order theory beyond the Gaussian approximation, and keep terms up to $A_4\phi^4/4!$ in $H_f$.  
The $\phi^4$ theory is valid if  the Taylor expansion of  $H_f[\zeta,\phi]$ 
can be truncated at the fourth order term for the dominant fluctuations $\phi$.
We therefore introduce the necessary condition for validity of the $\phi^4$ theory,
\begin{equation}
\label{criterion}
 \langle\phi({\bf r})^2\rangle<\frac{30A_4(\zeta)}{A_6(\zeta)}.
\end{equation}
If the criterion  (\ref{criterion})
is violated, the
$\phi^4$ theory may be oversimplified. Before discussing the results of our theory for particular models, 
we shall verify if the results for  $G({\bf r}_1,{\bf r}_2)$ satisfy (\ref{criterion}) for ${\bf r}_1={\bf r}_2$. 
 Note that the RHS of  (\ref{criterion}) depends on $\zeta$, therefore the accuracy of the $\phi^4$ 
theory can be different for different volume fractions. 

By construction of the mesoscopic theory,
$G({\bf r}_1,{\bf r}_2)$ is proportional to the correlation function for the microscopic density at
  the point belonging to a mesoscopic
region with the center at ${\bf r}_1$, and at the point belonging to a mesoscopic
region with the center at ${\bf r}_2$, averaged over these two mesoscopic regions~\cite{ciach:08:1,ciach:11:0}. 
In Ref.~\cite{ciach:08:1,ciach:11:0} it was assumed that these mesoscopic regions are smaller than the size of the aggragates,
and significantly larger than $\sigma$. 

In this work we limit ourselves to the disordered phase with 
the average volume fraction independent of the space position, $\bar\zeta=const$. 
In the disordered fluid 
 \begin{equation}
 \label{calG}
   {\cal G}= G({\bf r},{\bf r})
  \end{equation}
is a number indepenent of ${\bf r}$. It increases with increasing deviation from the average
 volume fraction in regions with excess density. It can be considered as a measure of the
inhomogeneity of the system, or compactness of the aggregates.  

Let us first consider the relation between $\bar \zeta$ and $\bar \mu$.
When $H_f$ is truncated at the $\phi^4$ term, then from (\ref{C1})
we obtain
\begin{equation}
\label{muhomm}
\beta \bar\mu\approx \bar\zeta\int d{\bf r}\beta V(r)+
A_1(\bar\zeta)+\frac{A_3(\bar\zeta)}{2}\langle\phi({\bf r})^2\rangle+\frac{A_4(\bar\zeta)}{3!}\langle\phi({\bf r})^3\rangle
+\frac{A_5(\bar\zeta)}{4!}\langle\phi({\bf r})^4\rangle.
\end{equation}
In order to evaluate $\bar\mu$, we need approximations for $\langle\phi({\bf r})^n\rangle$ with $n\le 4$.

 
Our approach is a generalization of 
the method developed in Ref.\cite{ciach:15:0}. We calculate the correlation function $G$ using the
OZ equation (\ref{CG}), and Eq.(\ref{C2}) for $C$. Since in the $\phi^4$ theory $G$ depends only on
$A_n$ with $n\le 4$ (see (\ref{Taylor2}), (\ref{Taylor2int}) and (\ref{Xav})),
in the consistent approach $C$ should be expressed in terms of   $A_n$ with $n\le 4$ too. Thus, from (\ref{C2})
we obtain for  the disordered phase
\begin{equation}
\label{selfcond}
 C({\bf r}_1,{\bf r}_2)\approx C^{(0)}_2({\bf r}_1,{\bf r}_2)+\frac{A_4(\bar\zeta)}{2}G({\bf r}_1,{\bf r}_2)\delta({\bf r}_1-{\bf r}_2)
 -\frac{A_3(\bar\zeta)^2}{4}\langle\phi({\bf r}_1)^2\phi({\bf r}_2)^2\rangle^{con}.
\end{equation}
In order to have a closed set of equations ((\ref{CG}) and (\ref{selfcond})), we need an approximation for
$\langle\phi({\bf r}_1)^2\phi({\bf r}_2)^2\rangle^{con}$. In
 Ref.\cite{ciach:15:0} we assumed $\langle\phi({\bf r}_1)^2\phi({\bf r}_2)^2\rangle^{con}=2G({\bf r}_1,{\bf r}_2)^2$, but
this approximation turns out to be an oversimplification for 3D systems, as discussed in sec.2. Thus, we must take into account
the well-known relation of the four-point correlation function with $G$ and the three- and 
four-point direct correlation functions~\cite{amit:84:0,evans:79:0}.
The expressions for $C_3$ and $C_4$ are obtained in our theory
from Eq.(\ref{Cn}). The approximate forms of $C_3$ and $C_4$ for $H_f$ Taylor-expanded up to $\phi^4$
(see (\ref{Taylor2})-(\ref{Taylor2int}))  are given in terms of the correlation functions in  Appendix A.
\begin{figure} 
\includegraphics[scale=0.34]{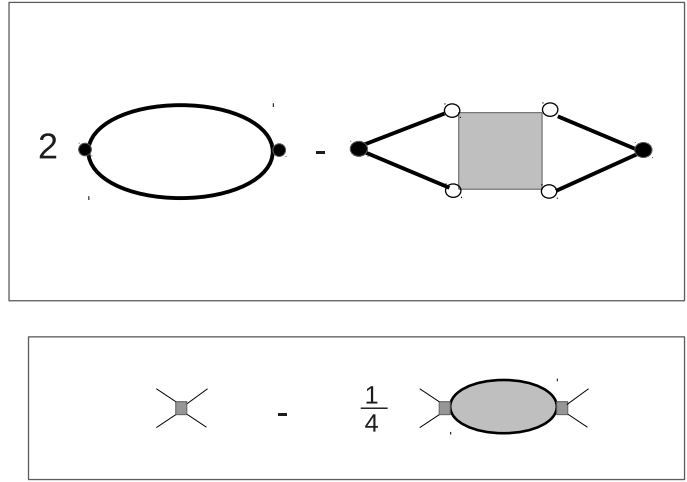} 
\begin{center}
 \caption{Upper panel: graphical representation of Eq.(\ref{G4}). 
 The  bullets represent the external points ${\bf r}_1$ and ${\bf r}_2$.
 The big shaded
 square represents  $C_4$. Open circles represent internal points. 
The thick line connecting points ${\bf r}'$ and ${\bf r}''$ represents $G({\bf r}',{\bf r}'')$. With each internal point ${\bf r}$, 
 an integration over ${\bf r}$ is associated. The first and second  diagram represents the  first and second  term in
  Eq.(\ref{G4}) respectively.
  Lower panel: graphical representation of  our approximation (\ref{C4app})  for the four-point direct correlation function $C_4$.
  The small shaded squares at the vertices represent $A_4$. Thin line emanating from the vertex at ${\bf r}_i$ 
  represents the corresponding argument of $C_4({\bf r}_1,{\bf r}_2,{\bf r}_3,{\bf r}_4)$. The first term is given in Eq.(\ref{C0n}).
  The shaded loop connecting the vertices at ${\bf r}'$ and  ${\bf r}''$
   is a solution of the self-consistent equation (\ref{Ds0}). It represents 
   our approximation for $\langle\phi({\bf r}')^2\phi({\bf r}'')^2\rangle^{con}$,  
  and is shown in terms of Feynman diagrams in the lower panel of Fig.\ref{Fd1}. }
 \label{G4fig}
  \end{center}
\end{figure}
In order
to obtain a closed set of equations for   $\langle\phi({\bf r}_1)^2\phi({\bf r}_2)^2\rangle^{con}$, we neglect 
the contribution proportional to $C_3$,
\begin{eqnarray}
\label{G4}
 \langle\phi({\bf r}_1)^2\phi({\bf r}_2)^2\rangle^{con}\approx 2 \langle\phi({\bf r}_1)\phi({\bf r}_2)\rangle^2
 \\\nonumber
 -\int d{\bf r}'\int d{\bf r}''\int d{\bf r}'''\int d{\bf r}'''' 
G({\bf r}_1,{\bf r}')G({\bf r}_1,{\bf r}''') G({\bf r}_2,{\bf r}'')G({\bf r}_2,{\bf r}'''')
C_4({\bf r}',{\bf r}'',{\bf r}''',{\bf r}''''),
\end{eqnarray}
 and for $C_4$ (Eq.(\ref{C4appendix}) in Appendix A) we make the approximation
\begin{eqnarray}
\label{C4app}
 C_4({\bf r}',{\bf r}'',{\bf r}''',{\bf r}'''') \approx
 A_4\delta({\bf r}'-{\bf r}'')\delta({\bf r}''-{\bf r}''')\delta({\bf r}'''-{\bf r}'''')
 \\\nonumber
 -\Big(\frac{A_4}{2}\Big)^2\Bigg[
 \langle\phi({\bf r}')^2\phi({\bf r}'')^2\rangle^{con} \big(\delta({\bf r}'-{\bf r}''')\delta({\bf r}''-{\bf r}'''')
 +\delta({\bf r}'-{\bf r}'''')\delta({\bf r}''-{\bf r}''')\big) 
 \\\nonumber+\langle\phi({\bf r}')^2\phi({\bf r}''')^2\rangle^{con}
 \delta({\bf r}'-{\bf r}'')\delta({\bf r}'''-{\bf r}'''')
 \Bigg].
\end{eqnarray}
Graphical representation of the above equations is shown in Fig.\ref{G4fig}.
We insert (\ref{C4app}) in (\ref{G4}),   and  obtain

\begin{eqnarray}
\label{g4a}
  \langle\phi({\bf r}_1)^2\phi({\bf r}_2)^2\rangle^{con}\approx 2 \langle\phi({\bf r}_1)\phi({\bf r}_2)\rangle^2
  \\\nonumber
  -A_4\int d{\bf r}'\langle\phi({\bf r}_1)\phi({\bf r}')\rangle^2\langle\phi({\bf r}')\phi({\bf r}_2)\rangle^2
   \\\nonumber
   +\Big(\frac{A_4}{2}
   \Big)^2\int d{\bf r}'\int d{\bf r}''\langle\phi({\bf r}_1)\phi({\bf r}')\rangle^2
   \langle\phi({\bf r}')^2\phi({\bf r}'')^2\rangle^{con}
   \langle\phi({\bf r}'')\phi({\bf r}_2)\rangle^2
   \\\nonumber
    +2\Big(\frac{A_4}{2}
   \Big)^2\int d{\bf r}'\int d{\bf r}''\langle\phi({\bf r}_1)\phi({\bf r}')\rangle\langle\phi({\bf r}_1)\phi({\bf r}'')\rangle
   \langle\phi({\bf r}')^2\phi({\bf r}'')^2\rangle^{con}\langle\phi({\bf r}')\phi({\bf r}_2)\rangle
   \langle\phi({\bf r}'')\phi({\bf r}_2)\rangle
\end{eqnarray}

\begin{figure}
   \includegraphics[scale=0.4]{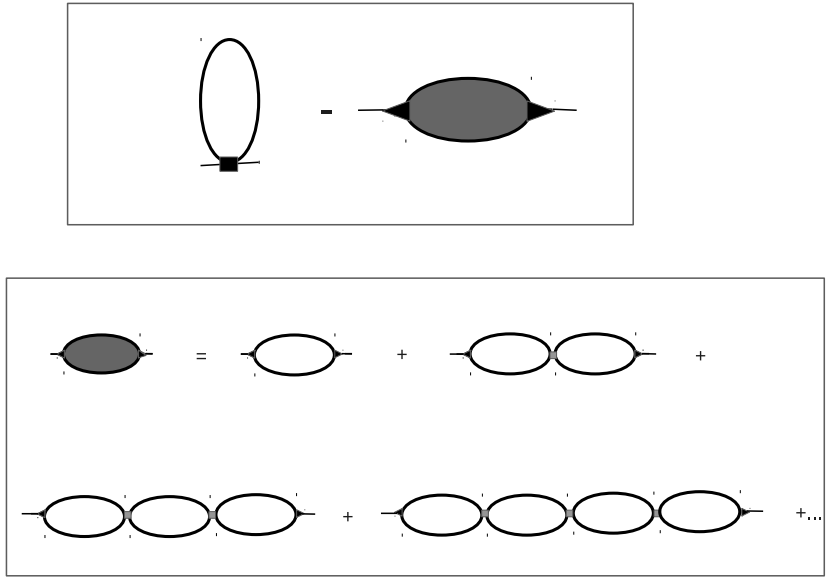}
   \begin{center}
\caption{Upper panel: diagrammatic representation of the fluctuation contribution to $C({\bf r}_1,{\bf r}_2)$. 
The triangles and squares represent $A_3$ and $A_4$
respectively.  Thin line emanating from a black vertex (triangle or square) at ${\bf r}_i$ 
represents the corresponding argument of  $C({\bf r}_1,{\bf r}_2)$.
The thick line connecting the points ${\bf r}'$ and ${\bf r}''$ represents $G({\bf r}',{\bf r}'')$. The shaded loop 
represents a series of diagrams shown in the lower panel.
 In the lower panel with each internal point at ${\bf r}$, 
shown as a gray square, an integration over ${\bf r}$ is associated. Finally,  a symmetry factor $1/2$ is associated with each loop.
In the Gaussian approximation only the first term in the series shown in the lower panel is included~\cite{ciach:15:0}.
}
\label{Fd1}
\end{center}
\end{figure}

When only the terms that have a form of
convolution are kept, we obtain from Eq.(\ref{g4a}) a very simple equation for 
$ \langle\phi({\bf r}_1)^2\phi({\bf r}_2)^2\rangle^{con}$
that in Fourier representation takes the form
\begin{eqnarray}
\label{Ds0}
 \tilde D_s(k)=2\tilde D(k) - A_4\tilde D(k)^2 + \Big(\frac{A_4}{2}\Big)^2\tilde D(k)^2 \tilde D_s(k),
\end{eqnarray}
where 
\begin{eqnarray}
\label{tildeD}
 \tilde D(k)=\int d({\bf r}_1-{\bf r}_2) e^{-i({\bf r}_1-{\bf r}_2)\cdot {\bf k}}G({\bf r}_1,{\bf r}_2)^2
\end{eqnarray}
 is  the Fourier fransform of  $G({\bf r}_1,{\bf r}_2)^2$, 
 and $\tilde D_s(k)$ is the Fourier transform of 
$ \langle\phi({\bf r}_1)^2\phi({\bf r}_2)^2\rangle^{con}$ defined in the same way. From (\ref{Ds0}) we easily get 
\begin{equation}
\label{Ds}
 \tilde D_s(k)= \frac{ \tilde D(k)}{1+\frac{1}{2}A_4\tilde D(k)}.
\end{equation}
%
In this approximation Eq.(\ref{selfcond}) in Fourier representation takes the simple form
\begin{equation}
\label{selfconm0}
\tilde C(k)=\tilde C^{(0)}_2(k)+\frac{A_4(\bar\zeta)}{2}{\cal G}-\frac{A_3^2(\bar\zeta)}{2}\tilde D_s(k).
\end{equation}
Eqs.(\ref{CG}), (\ref{selfconm0}),  (\ref{Ds}) and (\ref{tildeD}) form a closed set of equations for $G$. 
This result agrees with the results of the perturbation expansion in terms of Feynman diagrams in the self-consistent approximation, 
where only 1-loop diagrams and diagrams that have a form of chains of loops are included. The diagrams contributing 
to the direct correlation function $C$ in the approximation equivalent to our theory are shown in Fig.\ref{Fd1}.

Note that in contrast to the Gaussian approximation, positive solution of Eq.(\ref{selfconm0}) exists when $\tilde C^{(0)}_2(k)<0$
even for $\zeta\to 0$, since $\tilde D_s\ll\tilde D$, and the sum of the last two terms  on the RHS is positive and large. 
Recall that we have neglected the term of order $C_3^2G^5$ in Eq.(\ref{G4}). Since $C_3\propto A_3$, and $A_3(\zeta_c)=0$ 
where $\zeta_c$ is the critical density,
 the accuracy of the solution of Eq.(\ref{selfconm0}) 
decreases for increasing $|\zeta-\zeta_c|$. 

Let us return to the chemical potential, Eq.(\ref{muhomm}).
In the 
lowest-order approximation we keep the dominant terms only, i.e. we neglect 
$\langle\phi({\bf r})^3\rangle$ and assume
$\langle\phi({\bf r})^4\rangle\approx 3\langle\phi({\bf r})^2\rangle^2$. In this approximation 
\begin{equation}
\label{muhomma}
\beta \bar\mu\approx \bar\zeta\int d{\bf r}\beta V(r)+
A_1(\bar\zeta)+\frac{A_3(\bar\zeta)}{2}{\cal G}
+\frac{A_5(\bar\zeta)}{8}{\cal G}^2.
\end{equation}
The fluctuation contribution is given in the last two terms, where ${\cal G}$ must be obtained from the solution of
Eqs.(\ref{CG}), (\ref{selfconm0}),  (\ref{Ds}) and (\ref{tildeD}).
 Both $A_3(\zeta)$ and $A_5(\zeta)$ are negative for small-, 
and positive for large values of $\zeta$ (see (\ref{An})). Thus, the fluctuations lead to decreased and increased $\bar\mu$
for small and large $\zeta$
respectively, as found already in Ref.\cite{ciach:15:0} in the Gaussian approximation.
 Note that the leading-order corrections to (\ref{muhomma}), associated with $\langle\phi({\bf r})^3\rangle$
 and $\langle\phi({\bf r})^4\rangle-3\langle\phi({\bf r})^2\rangle^2$, are of the opposite sign to $A_{2n+1}$, hence 
 the difference between the exact and the MF result for $\bar\mu$ should be smaller than predicted by  (\ref{muhomma}).
 
The EOS is obtained from 
\begin{equation}
\label{p}
 p= -\Omega/{\cal V}=p^{MF}+k_BT\ln\Xi/{\cal V}
\end{equation}
where ${\cal V}$ denotes the system volume, $\Xi$ is defined in Eq.(\ref{Xi}), and 
$p^{MF}=-\Omega_{co}/{\cal V}$. From (\ref{Omco}) we have
\begin{eqnarray}
\label{pMF}
 p^{MF}=-\frac{1}{2}\tilde V(0)\bar \zeta^2 - f_h(\bar\zeta) +\bar\mu \bar\zeta .
\end{eqnarray}
Note that the fluctuation contribution to the pressure $p$ is included already in $p^{MF}$,
if for $\bar\mu$ we use our result (\ref{muhomma}) to obtain the EOS.
 
 We want to estimate the correction to $p^{MF}$ in the phase-space region where the MF predicts instability of the disordered phase
 with respect to periodic fluctuations. In this case  $C_2^{(0)}(k)<0$ for some range of $k>0$, 
 and the functional integrals in the standard perturbation expansion diverge. To overcome this problem 
 we write $\ln\Xi$ in the form
\begin{equation}
\label{lnXi}
 \ln\Xi= \ln\int D\phi e^{-\beta H_G}
 +\ln \Big(1+\langle \sum_{n=1}^\infty \frac{(-\beta\Delta H_G)^n}{n!}\rangle_G\Big)
\end{equation}
where $\langle ...\rangle_G$ is the average calculated with the probability proportional to $\exp(-\beta H_G)$, 
$H_f=H_G+\Delta H_G $, $H_G$ is given in (\ref{HG}), and 
\begin{eqnarray}
\label{DHG}
 \Delta H_G= -\frac{1}{2} \int d{\bf r}_1 d{\bf r}_2 \phi({\bf r}_1)\phi({\bf r}_2)
 \Big(\frac{A_4{\cal G}}{2} \delta({\bf r}_1 ,{\bf r}_2)
 -\frac{A_3^2}{2}\langle\phi({\bf r}_1)^2\phi({\bf r}_2)^2\rangle^{con}\Big)
 \\\nonumber
 +\int d{\bf r}\Big(C_1^{(0)}\phi({\bf r})+\frac{A_3}{3!}\phi({\bf r})^3+\frac{A_4}{4!}\phi({\bf r})^4\Big).
\end{eqnarray}
In the first term on the RHS of  Eq.(\ref{DHG}) the explicit form of $C_2^{(0)}-C$ is used (see(\ref{selfconm0})). 
We obtain the approximation for pressure
using  (\ref{lnXi}) and (\ref{DHG}), and  keeping only the leading-order contribution, 
 \begin{eqnarray}
\label{pa}
 p\approx p^{MF}+\frac{A_4{\cal G}^2}{8}.
\end{eqnarray}
 $k_BT \ln\int D\phi e^{-\beta H_G} /{\cal V}$  is disregarded in (\ref{pa}), since as shown in Ref.\cite{ciach:15:0}, this term
 is negligible compared to $\bar\mu \bar\zeta$.

The procedure developed in this section  is just the first step beyond the self-consistent Gaussian approximation.
Systematic improvement of the accuracy of the results is possible within
the framework described above, when $C_3$ and  the neglected terms of $C_4$  are taken into account in Eq.(\ref{G4}), 
and/or $H_f$ is truncated at 
a higher-order term. 
    
\section{results for the 1d lattice model with competing interactions}

In this section we verify the theory developed in sec.3  by comparing  the results with the exact solution of the 1d lattice model
with first-neighbor attraction $J_1$ and third neighbor repulsion $J_2$. 
For the comparison we choose $J=J_2/J_1=3$ \cite{pekalski:13:0}.
We shall verify if the present version of the theory yields  better agreement
with the exact results than the Gaussian approximation. 

Because of the repulsion between the third neighbors, clusters composed of three particles separated by three empty sites
are favourable energetically. Such an ordered periodic structure is stable only at $T=0$. For $T>0$ a disordered inhomogeneous phase
with oscillatory decay of correlations is stable. 
The characteristic properties of the disordered inhomogeneous phase 
determined for the considered model in Ref.\cite{pekalski:13:0} are: 
(i) the correlation length
increases rapidly to very large values for decreasing $T$ and $|\zeta-\zeta_c|$,  where $\zeta_c=1/2 $ 
  is the volume fraction optimal for the periodic distribution of clusters in this model (it is also equal to the critical volume fraction),
  (ii) compressibility is
very small for the volume fraction optimal for the periodic pattern, $\zeta\approx 1/2$,
and decreases significantly with decreasing $T$, (iii) the compressibility is very large 
for small and large $\zeta$, and increases with decreasing $T$, (iv) the $\bar\mu(\zeta)$ line has three inflection points.
Many of the above features are predicted by the Gaussian theory~\cite{ciach:15:0}.
However, in the Gaussian approximation the compressibility  for $\zeta\approx 1/2$ is independent of $T$, and no anomalous decrease of
pressure in a heated system is obtained for $\zeta\approx 0.6$, in contrast to the exact results.

Note that when $\bar\zeta=const.$ the only difference between Eq.(\ref{selfconm0}) and Eq.(\ref{selfcon}) 
(rewritten in Fourier representation) is the replacement
of $\tilde D$ by $\tilde D_s$ (see (\ref{tildeD}) and (\ref{Ds})).
Thus, to find $C$
we can repeat the procedure described in detail in Ref.\cite{ciach:15:0}. In short, for $ \tilde C^{(0)}_2$ defined in Eq.(\ref{C02}) 
we make the approximation
\begin{equation}
\label{tildeCa}
 \tilde C^{(0)}_2(k)\approx c_0 +v_0(k^2-k_0^2)^2.
\end{equation} 
 The parameters  $c_0,v_0,k_0$ are obtained from the form of the interaction potential
and from the form of $f_h$ in the lattice-gas model by fitting (\ref{tildeCa}) to (\ref{C02}) 
for $k$ close to $k_0$ corresponding to the minimum
of  $ \tilde C^{(0)}_2$~\cite{ciach:15:0}. For $\tilde C$
we postulate the same form i.e. Eq.(\ref{tildeCa}), but with the parameters $c_0,v_0,k_0$ 
replaced by the renormalized ones, $c_r,v_r,k_r$, respectively. The equations for $c_r,v_r,k_r$ 
are analogous to the  equations (39)-(41) in Ref.\cite{ciach:15:0}, but with $\tilde D(k)$ replaced 
by $\tilde D_s(k)$. For the form of $\tilde D(k)$  and more details
see~Ref.\cite{ciach:15:0}. 

\begin{figure}
\begin{center}
  \includegraphics[scale=0.3]{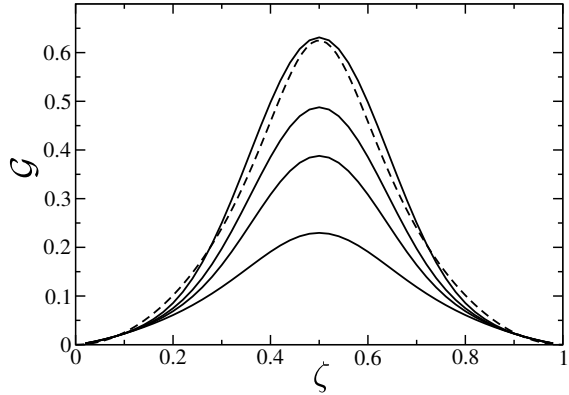}
  \caption{ ${\cal G}$ describing the local excess of the volume fraction over the average value
  for the 1D lattice model with the repulsion to attraction ratio $J=3$
  for $T^*=0.5, 0.6, 0.7, 1$ (solid lines, from top to bottom), and the RHS of Eq.(\ref{criterion}) (dashed line).
  When the dashed line lies below the solid line, the $\phi^4$ theory may be oversimplified. }
 \label{G01d}
 \end{center}
  \end{figure}

We solve the  equations for $c_r,v_r,k_r$ numerically in this  part of the phase space, 
where MF predicts instability of the disordered phase.
 The MF line of instability, obtained from $\tilde C_2^{(0)}(k_0)=0$, is  $k_BT=-\tilde V(k_0)/A_2(\zeta)$, and the interesting 
 thermodynamic states are $k_BT<-\tilde V(k_0)/A_2(\zeta)$. For such states inhomogeneous distribution of particles is more
 probable than the homogeneous one.
 Following Ref.\cite{pekalski:13:0,ciach:15:0}, we introduce dimensionless  temperature, 
 $T^*=k_BT/J_1$, and consider $T^*\le 1$, where strong inhomogeneities are predicted by the exact results.
 We first verify if the necessary condition for validity of the $\phi^4$ theory,  Eq.(\ref{criterion}), is satisfied for 
  $T^*\le 1$.
In Fig.\ref{G01d} we compare ${\cal G}$  (see (\ref{calG}))
with the RHS of Eq.(\ref{criterion}) for several temperatures. 
  We can see that for all the temperatures the $\phi^4$ theory is  oversimplified  for $\zeta<0.1$ and $\zeta>0.9$.
  Moreover, for $T^*\le 0.5$ the criterion (\ref{criterion}) is violated for $0.25<\zeta<0.75$. Thus, we should 
  limit ourselves to $T^*>0.5$ and $0.1<\zeta<0.9$. The case $T^*\le 0.5$ will be shown to see how 
  the oversimplified theory compares with the exact results.
  
\begin{figure}
  \includegraphics[scale=0.3]{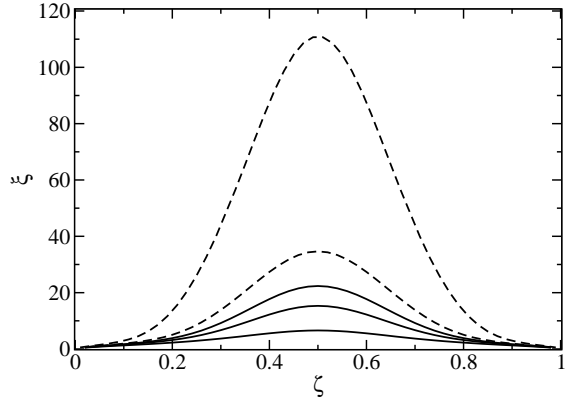}
 \begin{center}
  \caption{The  correlation length for the 1D lattice model with the repulsion to attraction ratio $J=3$
  for $T^*=0.3, 0.5, 0.6, 0.7, 1$ (from top to bottom). Dashed lines correspond to $T^*$ for 
  which the criterion (\ref{criterion}) is not satisfied
  and the $\phi^4$ theory is oversimplified.}
  \label{xi1d}
   \end{center}
  \end{figure}

   In Fig.\ref{xi1d} the correlation length is presented for a few temperatures.
  We can see that the  correlation length  $\xi$ increases with decreasing temperature and/or $|\zeta-\zeta_c|$. 
  This behavior agrees  with the exact results. However,  when the necessary condition
  (\ref{criterion}) is not satisfied and our  $\phi^4$ theory is oversimplified, 
  we obtain significantly smaller $\xi$ than found in Ref.\cite{pekalski:13:0}. 
  The difference between our predictions and 
  the exact results increases for decreasing $T^*$,  
  because  the accuracy of our theory decreases 
  with decreasing temperature (see Fig.\ref{G01d}). 
  
\begin{figure}
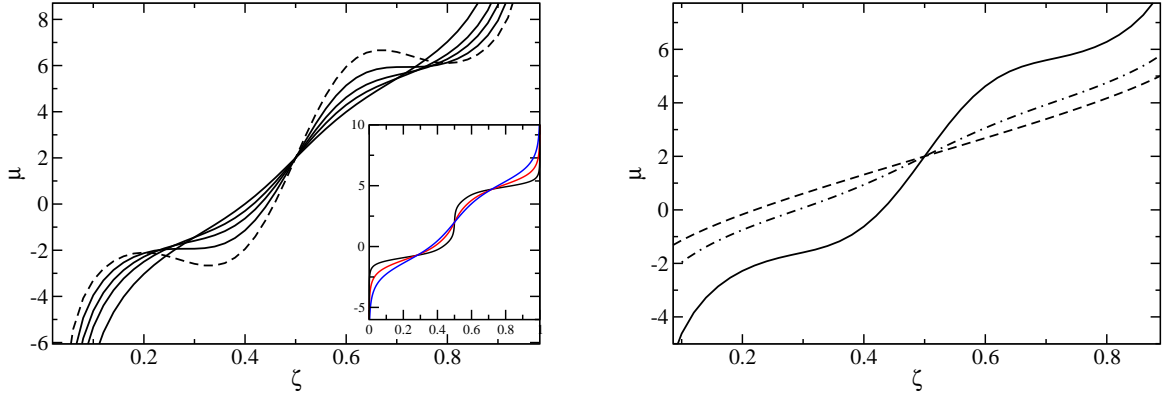

   \begin{center}
   \includegraphics[scale=0.3]{fig5a.eps}
   \hskip1cm	
  \includegraphics[scale=0.3]{fig5b.eps}
  \caption{The chemical potential $\mu$  in our theory (see Eq.(\ref{muhomma})) for the 1D lattice model with the repulsion to 
  attraction ratio $J=3$ in $k_BT$ units (a)
  for $T^*=0.5$ (dashed line) and $T^*= 0.6, 0.7, 0.8, 1$ (solid lines, from top to bottom on the left). Note that for $T^*=0.5$
  the $\phi^4$ theory is oversimplified. In the inset the exact results obtained in Ref.~\cite{pekalski:13:0} for
  $T^*=0.4,0.7,1$ (top to bottom line on the left) are shown.(b) $\mu$
  for $T^*=0.7$   in our theory (solid line), in MF  (dashed line), and in the Gaussian approximation (dash-dotted line). }
 \label{mu1d} 
 \end{center}
 \end{figure}

  In Fig.\ref{mu1d}a we present the chemical potential for a few temperatures, 
  and  in the inset the exact results are shown for comparison.
  In Fig.\ref{mu1d}b
  predictions of  our theory are compared with the results of the  MF and Gaussian theories for $T^*=0.7$. 
  By the MF prediction we mean here Eq.(\ref{muhomma}) without the fluctuation contributions (the last two terms), but in MF the disordered phase 
  is unstable with respect to periodic ordering for the considered temperatures.
  \begin{figure}
  \begin{center}
  \includegraphics[scale=0.3]{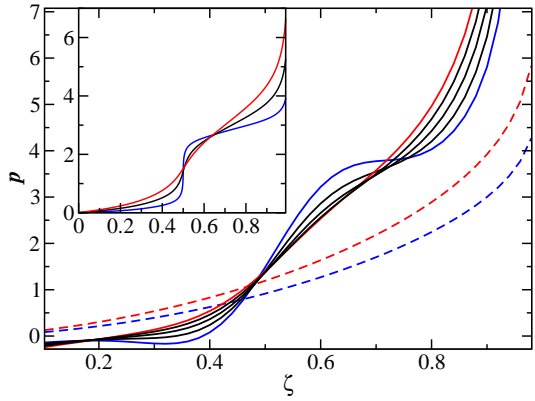}
  \caption{The pressure isotherms for the 1D lattice model with the repulsion to attraction ratio $J=3$ in $k_BT$ units
  for $T^*=0.6, 0.7, 0.8,0.9, 1$ (solid lines, from  bottom to top on the right). 
  The dashed lines show the MF result for $T^*=0.6  ,1$. In the inset the exact results for $T^*=0.4,0.7,1$ 
  (bottom to top line on the right) are shown. Note the corret prediction of the anomalous increase of pressure for 
  decreasing temperature for a range of volume fractions too large for the optimal distribution of the clusters, $0.5<\zeta<0.7$.}
  \label{p1d} 
  \end{center}
  \end{figure}
   Fig.\ref{mu1d}a suggests  mechanical instabilities for two intervals of $\zeta$ at $T^*=0.5$. 
   The exact results for very low $T^*$, however,
   show only very large 
  compressibility  and pseudo-phase transitions (very large change of $\zeta$ for very small
  change of $\bar\mu$)
  at $\zeta\approx 0.25, 0.75$. As can be seen in Fig.\ref{G01d}, for $T^*=0.5$
   the $\phi^4$ theory is oversimplified. Thus, these instabilities 
  are artifacts resulting from the truncation of the Taylor expansion of $H_f$ at the $\phi^4$ term. 
  Very large compressibility for large volume fractions usually signals an approach to a phase transition. In this model
  there are no phase transitions for $T^*>0$, but the properties of the disordered phase differ significantly from the properties 
  of the disordered phase in simple fluids.
  We conclude that peculiar behavior, such as the pseudo phase transitions observed in Ref.\cite{pekalski:13:0},
  but not necessarily a real phase transition should be expected when mechanical instabilities
  are predicted by our approximate theory, especially when the criterion (\ref{criterion}) is not satisfied. 
  
  From Fig.\ref{mu1d} we can  see  significant improvement of our approximation compared to the Gaussian theory~\cite{ciach:15:0} for $T^*>0.5$. 
   Our theory and the exact results  both 
 indicate that for increasing $\zeta$ the slope of the $\mu(\zeta)$ line is  small (large compressibility) for $\zeta\sim 0.25$, 
 then  increases to much larger values for $\zeta\approx\zeta_c$
 (the compressibility decreases),
 and decreases again for $\zeta\approx 0.75$.
 The crossover from large to small to large compressibility occurs more and more rapidly when the temperature decreases, 
 and the sequence of very large - very small - very large
 compressibility is obtained for $0.2<\zeta<0.8$ at low $T^*$. 
  The slope of the $\mu(\zeta)$ line at $\zeta=0.5$ increases with decreasing $T^*$, and 
  the lines corresponding to different $T^*$ 
  intersect in three points, $\zeta\approx 0.25,0.5,0.75$, in agreement with the exact results. 
  The value of $\mu$ at 
  the points of intersection, $\mu\approx-2, 2, 6$, is only in  semiquantitative agreement with  $\mu=-2/3, 2, 14/3$ 
  obtained exactly. The accuracy of the present approximation decreases for increasing $|\zeta-\zeta_c|$ 
  (see Fig.\ref{mu1d} and \ref{p1d}). 
  This should be expected, because as discussed below Eq.(\ref{selfconm0}),
  for increasing $|\zeta-\zeta_c|$ the neglected contribution associated with $C_3$ in Eq.(\ref{G4}) increases, therefore 
   the smaller is  $|\zeta-\zeta_c|$, the better is the accuracy of the present approximation.
  
  We conclude that the effects of fluctuations in our approximation are overestimated, but
  as discussed in sec.3, we expect that  better approximation
  for $\langle\phi({\bf r}_1)^2\phi({\bf r}_2)^2\rangle^{con}$, and
  the  higher-order corrections to $\mu$, should lead to smaller deviations of $\mu$
  from $\mu^{MF}$, hence to better agreement with the exact results.
  The conseqence of the too small value 
  of $\mu$ for $\zeta<0.5$ is the negative value of pressure, as can be seen in Fig.\ref{p1d}. 
  Despite the negative values for small $\zeta$,
  the shape of the $p(\zeta)$ lines agrees quite well with the exact results~\cite{pekalski:13:0}. 
  In particular, we obtain the anomalous decrease of pressure for increasing $T$ for $\zeta\approx 0.6$.
  
 \section{Results for a  3d SALR model}

\begin{figure}
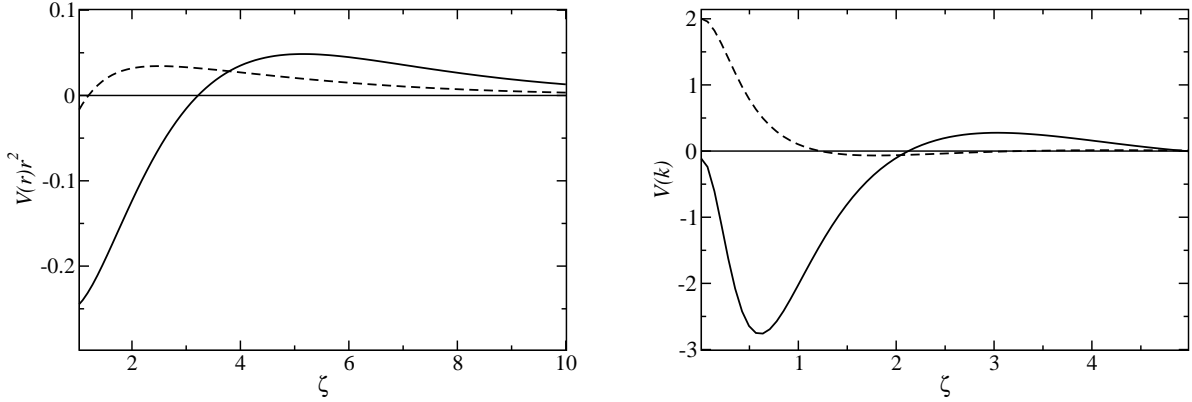

 \includegraphics[scale=0.3]{fig7a.eps}\hskip1cm
 \includegraphics[scale=0.3]{fig7b.eps}
 \caption{ Left: $V(r)r^2$ with $V(r)$ defined in (\ref{int_pot_r}) for System 1 (dashed line) and System 2 (solid line). Right:
 $\tilde V(k)$ (Eq.(\ref{uLRIIF})) for System 1 (dashed line) and System 2 (solid line) }
 \label{V}
\end{figure}

 We consider charged particles with hard cores of diameter $\sigma$ taken as a length unit.
 The particles repel each other at large distances with screened electrostatic interactions,
 and attract each other at short distances with solvent-mediated effective potential.
 The reference-system  free energy is given by the Percus-Yevick approximation
 \begin{eqnarray}
\label{PY}
\beta f_h(\zeta)=\rho^*\ln(\rho^*)-\rho^*+
\rho^*\Bigg[\frac{3\zeta(2-\zeta)}{2(1-\zeta)^2}-\ln(1-\zeta)\Bigg],
\end{eqnarray}
 where  $\rho^*= 6\zeta/\pi$ is the dimensionless density.
 For the interaction potential we choose the form 
 studied in Ref.\cite{archer:07:0,toledano:09:0,ciach:10:1,ciach:12:0,ciach:13:0}
 \begin{equation}
\label{int_pot_r} 
V(r)=\Big[-\frac{{\cal
A}_1}{r}e^{-z_1r}+\frac{{\cal A}_2}{r}e^{-z_2r}\Big]\theta(r-1),
\end{equation}
where $r$ is in  $\sigma$  units. 
In  Fourier representation $V$ takes the form
\begin{equation}
\label{uLRIIF}
\tilde V(k)=4\pi\Bigg[\frac{{\cal A}_2e^{-z_2}}{z_2^2+k^2}
\Big(z_2\frac{\sin k}{k}+\cos k\Big)-\frac{{\cal A}_1e^{-z_1}}{z_1^2+k^2}
\Big(z_1\frac{\sin k}{k}+\cos k\Big)\Bigg].
\end{equation}
$\tilde V(k)$ represents the increase of the system energy per unit volume when a volume-fraction 
wave with the wavenumber $k$ and unit amplitude is excited in the initially homogeneous system. 
We  choose two sets of parameters, considered in
Ref.~\cite{ciach:10:1} in the context of the most probable inhomogeneous
structures and in Ref.\cite{ciach:12:0} in the Gaussian approximation of the Brazovskii
type, i.e. in a linear order in the parameters $A_n$,
\begin{eqnarray}
\label{parameters}
System~1: && \hskip1cm {\cal A}_1=1,\quad {\cal A}_2=0.05, \quad z_1=3, \quad z_2=0.5;
\nonumber  \\
System~2: && \hskip1cm {\cal A}_1=1, \quad {\cal A}_2=0.2, \,\,\, \quad z_1=1, \quad z_2=0.5.
\end{eqnarray}
 $V(r)r^2$ for both systems is shown in Fig.\ref{V}a and $\tilde V(k)$ is shown in Fig.\ref{V}b.
 In both systems the global minimum of $\tilde V(k)$ is assumed for $k_0>0$, with  $k_0\approx 1.8$ in
System 1, and  $k_0\approx 0.6$ in System 2, and $\tilde V(k_0)<0$. Thus, the volume-fraction wave with the wavelength $2\pi/k_0$ 
is more probable than the homogeneous distribution of the particles. 

In System 1 the attraction strength and range are small, and the repulsion dominates, but it is not very strong either.
 Separation into dilute and dense homogeneous phases is neither entropically nor energetically
favourable when $\tilde V(0)>0$.  Because  $\tilde V(0)>0$ in System 1, the phase separation is 
 less favourable than the homogeneous distribution of particles. The latter is in turn energetically
 less favourable than formation of small compact clusters (presumably of thetrahedral shape).
 Since the minimum of $\tilde V(k)$ is shallow,  the energy gain associated with density waves
 with the wavelengths somewhat different from $2\pi/k_0$  is comparable. Such  waves  
can be excited with quite high probability. Thus,  small clusters 
at different separations for different $\zeta$, rather than transitions between an ordered phase with the period $2\pi/k_0$
and homogeneous dilute or dense phases should be expected.   

In System 2 the strength and range of attraction are both much larger than in System 1, 
and clusters larger   than in System 1 are formed.
The repulsion is  stronger too, and
the repulsive and attractive parts of the potential compete.
The global minimum of $\tilde V(k)$ at $k_0$ is deep, and the density waves with the wavenumber $k_0$ are
energetically favored over waves with different wavenumbers more strongly than in System 1.
Thus, we can expect stronger tendency for periodic
order with the period $2\pi/k_0$.  
Since $\tilde V(0)<0$, the phase separation is energetically favoured over the homogeneous state.  
When the average volume fraction of particles does not allow for the preferable periodic structure, the phase separation 
might compete with the periodic ordering. 

In MF periodic ordering of clusters or voids 
into lamellar, hexagonal, gyroid and bcc structurs was found for both systems~\cite{ciach:10:1}. 
It is well known, however that the periodic order is destroyed by fluctuations for a large part of the MF stability 
region of the ordered phases \cite{almarza:14:0,brazovskii:75:0}. Thus, we are interested in the phase-space region below 
the MF boundary of stability of the disordered phase, $T^*<1/A_2(\zeta)$, where $T^*=k_BT/|\tilde V(k_0)|$  
is the dimensionless  temperature introduced in Refs.\cite{ciach:08:1,ciach:10:1}. We want to find out
how the effects of fluctuations depend
on the range and strength of the attractive and repulsive parts of the SALR potential. 
Since 
$1/A_2(\zeta)\le 1/A_2(\zeta_c)\approx 0.024$, where $\zeta_c$ is the critical volume fraction \cite{ciach:12:0}, 
we shall consider $T^*<0.02$. Note that  the temperature scales in the 1D and 3D models, introduced
in Ref.\cite{pekalski:13:0} and Ref.\cite{ciach:10:1} respectively, are different.

We make the same approximation (\ref{tildeCa}) for $\tilde C_2^{(0)}$ as in the 1D case.  With this assumption,
$\tilde G_2^{(0)}(k)=1/\tilde C_2^{(0)}(k)$ in real-space representation has the form
\begin{equation}
\label{G3d}
 G^{(0)}(r)=A_0 e^{-r/\xi_0} \frac{\sin(\alpha_0 r)}{r},
\end{equation}
where the parameters are given in terms of $ c_0, v_0, k_0$  in Appendix B.
We postulate that $\tilde C(k)=c_r+v_r(k^2-k_r^2)^2$. 
With the above assumption  $G$  in real-space representation is given by Eq.(\ref{G3d}),
but with $\xi_0,\alpha_0,A_0$
replaced by $\xi_r,\alpha_r,A_r$ expressed in terms of $c_r,v_r,k_r$ by equations analogous
to Eqs.(\ref{A0})-(\ref{xi0}) in Appendix B. The expression for $\tilde D(k)$ (see (\ref{tildeD})) can be easily obtained,
and is also given in Appendix B.
We solve numerically the equations for $c_r,v_r,k_r$ that follow from
Eqs.(\ref{CG}), (\ref{selfconm0}),  (\ref{Ds}) and (\ref{tildeD}), using the procedure described in sec.3,
and in more detail in Ref.\cite{ciach:15:0}. 

In fact $\tilde C$ should be determined by solving 
directly the integral equations (\ref{CG}), (\ref{selfconm0}),  (\ref{Ds}) and (\ref{tildeD}), 
because the assumed functional
form of $\tilde C$ is appropriate only for systems with strong mesoscopic inhomogeneities and isotropic correlations.
 As shown in Ref.\cite{almarza:14:0,barci:13:0}, anisotropic correlations may appear in systems with competing interactions.
 Moreover, as argued in Ref.\cite{ciach:15:0}, Eq.(\ref{tildeCa}) can be a fair approximation when $\tilde V(k)$ 
 assumes a deep minimum.  For System 1 this approximation 
 can be too crude. 
Since solving the full equations is a very difficult task, we have decided to make all the above simplifying assumptions 
to obtain preliminary results.


 \subsection{System 1}
 
\begin{figure}
\begin{center}
  \includegraphics[scale=0.3]{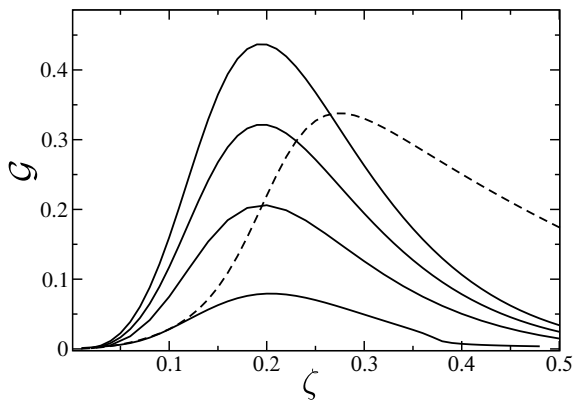}
 \caption{ ${\cal G}$ (Eq.(\ref{calG})) describing local deviations from the average volume fraction for  System 1,  
 as a function of the volume fraction $\zeta$
 for $T^*=0.0015, 0.002, 0.003, 0.015$ (solid lines, from  top to  bottom), and the RHS of Eq.(\ref{criterion}) (dashed line). 
 When the dashed line lies below the solid line, the $\phi^4$ theory may be oversimplified.}
 \label{G0S1}
\end{center}
\end{figure}

 Let us first  verify if the 
 necessary condition (\ref{criterion}) for validity of the $\phi^4$ theory is satisfied for the interesting temperature range.
 In Fig.~\ref{G0S1}  ${\cal G}$ 
 is shown together with $30A_4/A_6$ calculated for the Percus-Yevick approximation (\ref{PY}), with $A_n$ defined in (\ref{An}). 
 We can see that in the absence of the particle-hole symmetry  the necessary condition (\ref{criterion})
is and  is not satisfied  for large and for small  volume fractions respectively.
 The accuracy of the approximation increases 
 with increasing temperature, when the  inhomogeneities measured by ${\cal G}$ decrease. The necessary condition for 
 validity of the $\phi^4$ theory 
 is  satisfied for the whole range of $\zeta$ when  $T^*>0.015$.

\begin{figure}
\begin{center}
 \includegraphics[scale=0.3]{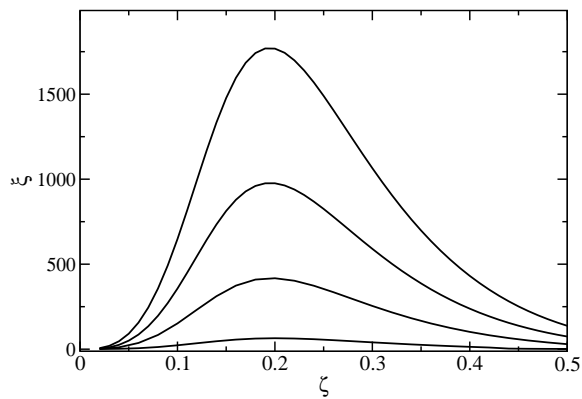}
 \caption{The correlation length $\xi$ for System 1 in $\sigma$-units as a function of the volume fraction.
  $T^*=0.0015, 0.002, 0.003, 0.007$ (solid lines, from  top to  bottom)}
 \label{xiS1}
 \end{center}
\end{figure}

\begin{figure}
\begin{center}
 \includegraphics[scale=0.27]{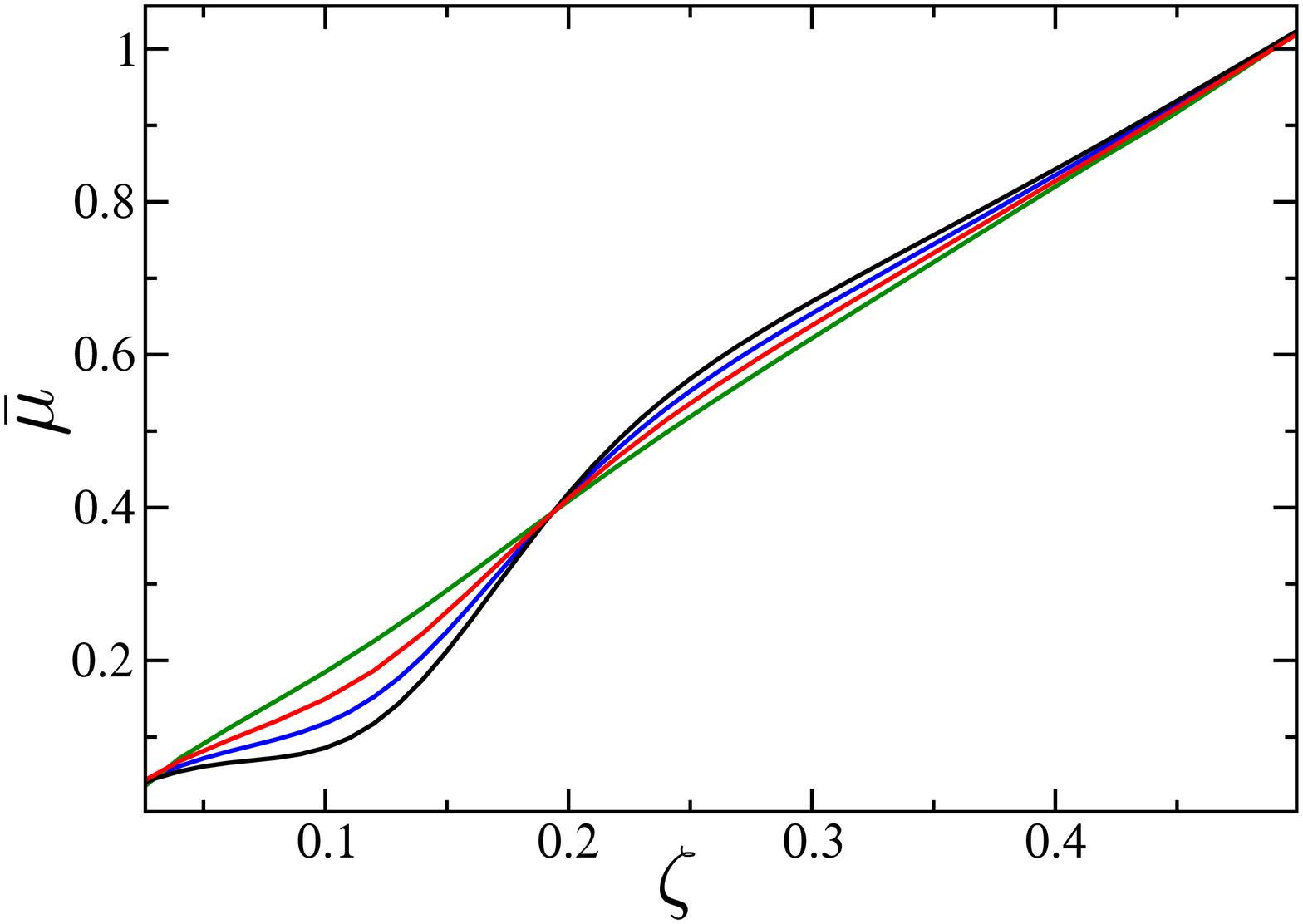}
 \hskip0.5cm
 \includegraphics[scale=0.27]{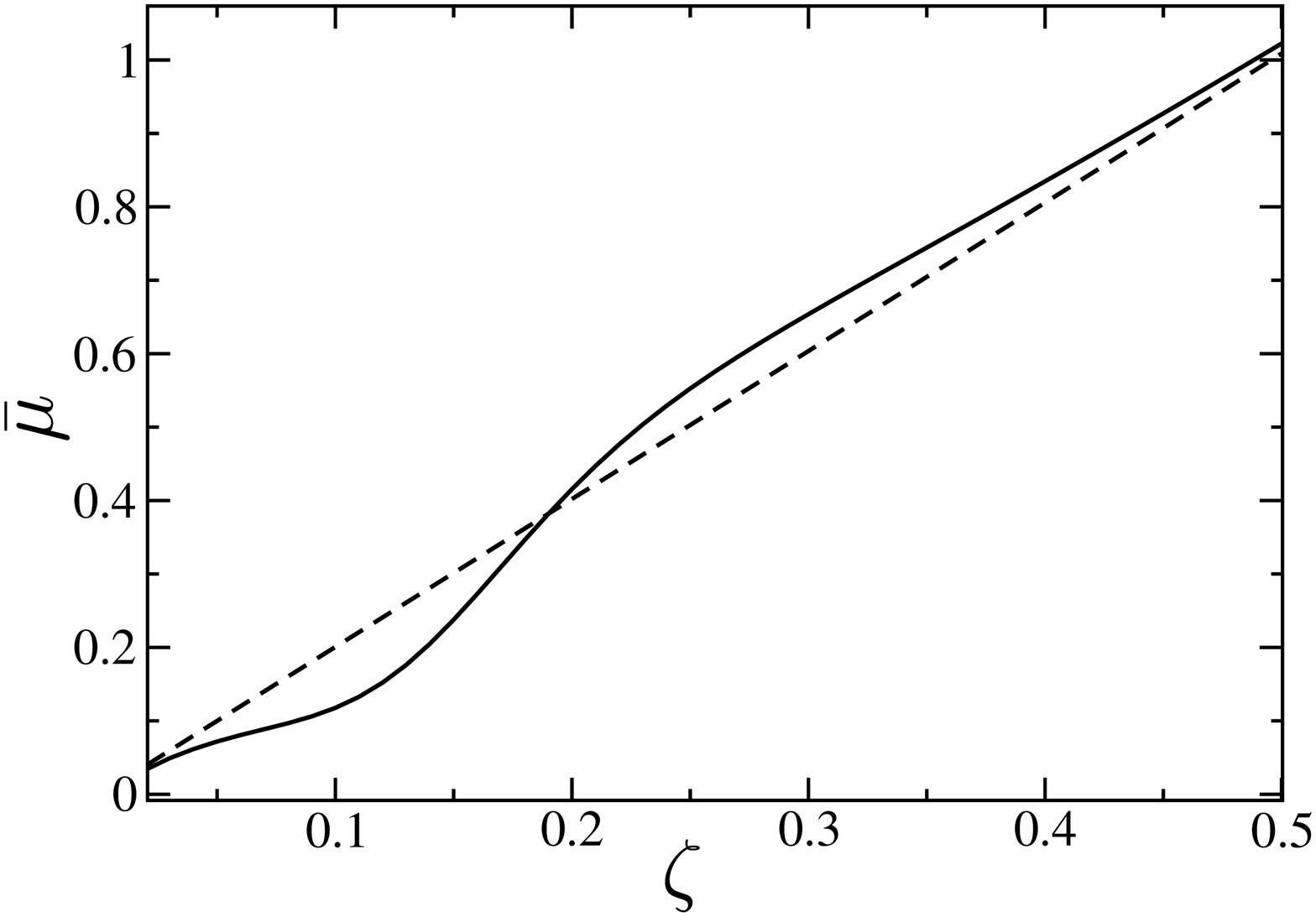}
 \caption{The chemical potential isotherms  in $k_BT$ units in our theory (see Eq.(\ref{muhomma})) for  System 1
 as a function of the volume fraction. (a) $T^*=0.0015, 0.002, 0.003, 0.007$ (solid lines, from  top to  bottom on the right).
 b) The chemical potential in our theory 
 (solid line) and in MF approximation, i.e. with the last two terms in Eq.(\ref{muhomma}) neglected (dashed line) for $T^*=0.002$. }
 \label{muS1}
\end{center}
\end{figure}

\begin{figure}
\begin{center}
 \includegraphics[scale=0.3]{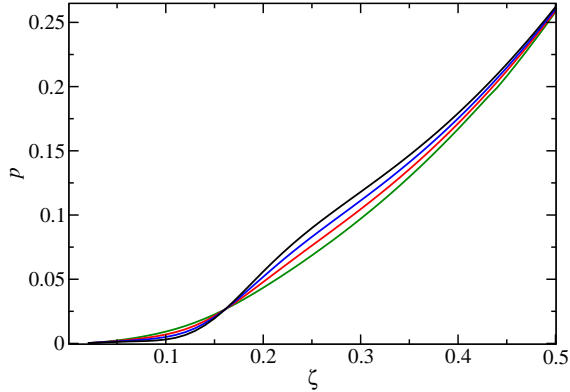}
 \caption{The pressure isotherms for  System 1  in $k_BT/\sigma^3$ units as a function of the volume fraction.
 $T^*=0.0015, 0.002, 0.003, 0.007$  from  top to  bottom line on the right. 
 Note the anomalous increase of pressure for decreasing $T$ for $\zeta>\zeta_c$, 
 similar to the anomaly obtained exactly in Ref.\cite{pekalski:13:0} for the 1D model.}
 \label{pS1}
 \end{center}
\end{figure}

 The correlation length, the chemical potential and pressure are shown in Figs.\ref{xiS1}-\ref{pS1}.
  Note a very large correlation length for $T^*\le 0.003$.   At $T^*=0.0015$ the
 necessary condition (\ref{criterion}) is satisfied  for $\zeta>0.25$, 
 and we can see that $\xi\sim 10^3$ for volume fractions $\zeta\sim 0.3$. The oscillatory decay of correlations
 with the mesoscopic period and the very large correlation langth is a signature of the mesoscopic inhomogeneity for the corresponding
 thermodynamic parameters. 

 The results for $\bar\mu(\zeta)$ and $p(\zeta)$ show large slope (indicating small compressibility) for $\zeta\approx\zeta_c$.
 The slope of  $\bar\mu(\zeta)$ and $p(\zeta)$ increases  with decreasing $T^*$ (the compressibility decreases).
 The  slope  of   $\bar\mu(\zeta)$  and $p(\zeta)$ is small 
 (large  compressibility) for $\zeta\approx 0.1$ 
 and  decreases with decreasing $T^*$ (the compressibility increases). For $\zeta$ increasing from $\zeta\approx 0.2$ the 
slope of   $\bar\mu(\zeta)$ and $p(\zeta)$  decreases a little, but neither 
increasing $\zeta$ nor decreasing $T^*$ leads 
 to very large compressibility in this system.
 Even for very small $T^*$, where our theory is expected to overestimate the effects of mesoscopic fluctuations, 
 the compressibility at
 $\zeta_c$ is significantly larger, and for $\zeta\sim 0.3$ the compressibility is
 significantly smaller than in the 1D system with strong repulsion. Note that  because
 the repulsion barrier is small,
 the increase of the volume fraction for $\zeta\approx\zeta_c$ does not require very large increase of $\bar\mu$ or $p$, 
 therefore the compressibility is not very small. 
 Our results show no sign
 of the phase transition or pseudo phase transition between the inhomogeneous
 and the dense homogeneous phase
 for $\zeta>\zeta_c$  for the considered range of $T^*$. Recall that in System 1 the attraction range is very small and
 $\tilde V(0)>0$. For this reason if the
 volume fraction is too large for formation of the 
 periodic structure with the wavenumber $k_0$, instead of the (pseudo)phase transition to the dense phase
 a decrease of the separation between the small clusters takes place.
 We cannot predict if 
 a phase transition (or a pseudo phase transition found for the 1D model in Ref.\cite{pekalski:13:0})
 between dilute gas and  inhomogeneous fluid  can occur for  $\zeta<\zeta_c$, because our $\phi^4$ theory 
 is oversimplified for  $\zeta<\zeta_c$ and $T^*<0.015$. 

 \subsection{System 2}
 
\begin{figure}
\begin{center}
 \includegraphics[scale=0.3]{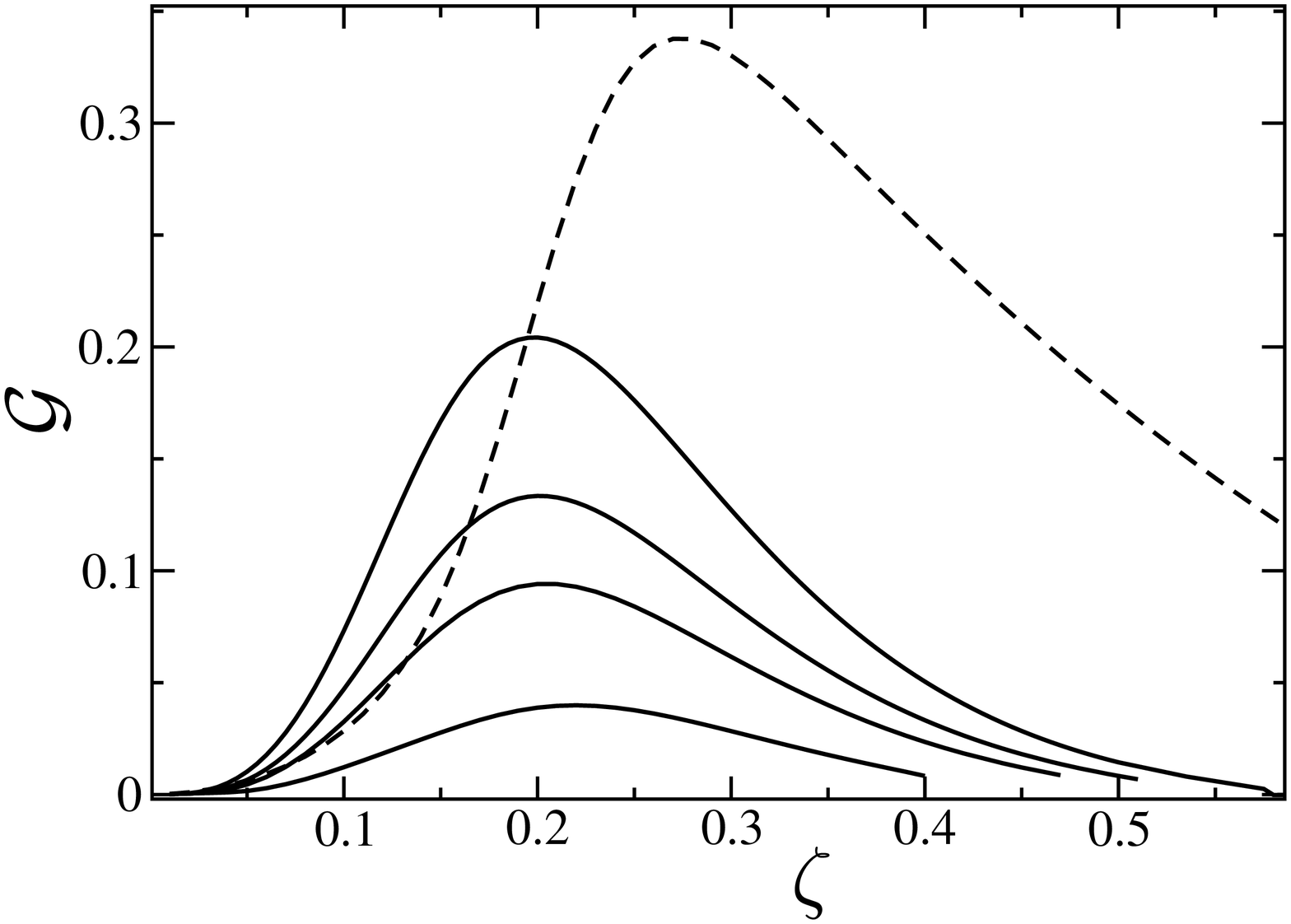}
 \caption{ ${\cal G}$ (Eq.(\ref{calG})) describing local deviations from the average volume fraction for  System 2  as a function of the volume fraction for
 $T^*=0.005,0.007,0.009.0.015$ (solid lines, from top to bottom), and the RHS of Eq.(\ref{criterion}) (dashed line).
 When the dashed line lies below the solid line, the $\phi^4$ theory may be oversimplified.}
 \label{G0S2}
\end{center}
\end{figure}

\begin{figure}
\begin{center}
 \includegraphics[scale=0.3]{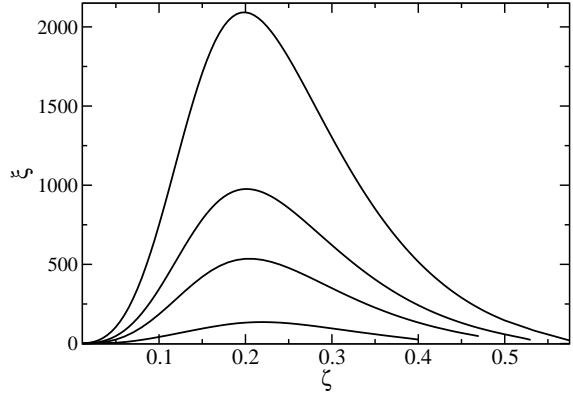}
 \caption{Correlation length for System 2 in $\sigma$-units as a function of the volume fraction.
 $T^*=0.005,0.007,0.009.0.015$ (solid lines, from top to bottom).}
 \label{xiS2}
 \end{center}
\end{figure}

 
 We first verify for which thermodynamic states our $\phi^4$ theory is not oversimplified.
 As shown in Fig.\ref{G0S2}, we obtain very similar behavior of ${\cal G}$ as 
 in System 1, namely the $\phi^4$ theory is oversimplified for small 
 $\zeta$ whose range increases with decreasing temperature. For $T^*\ge 0.015$,
 however, the necessary condition (\ref{criterion}) is satisfied for the whole range of $\zeta$.
The correlation length is  larger than in System 1 for the same dimensionless temperature,
as can be seen by comparison of Figs.\ref{xiS1}  and \ref{xiS2}. Thus, the tendency for periodic order is stronger, as expected
based on the interaction potentials (Fig.\ref{V}).

\begin{figure}
\begin{center}
  \includegraphics[scale=0.3]{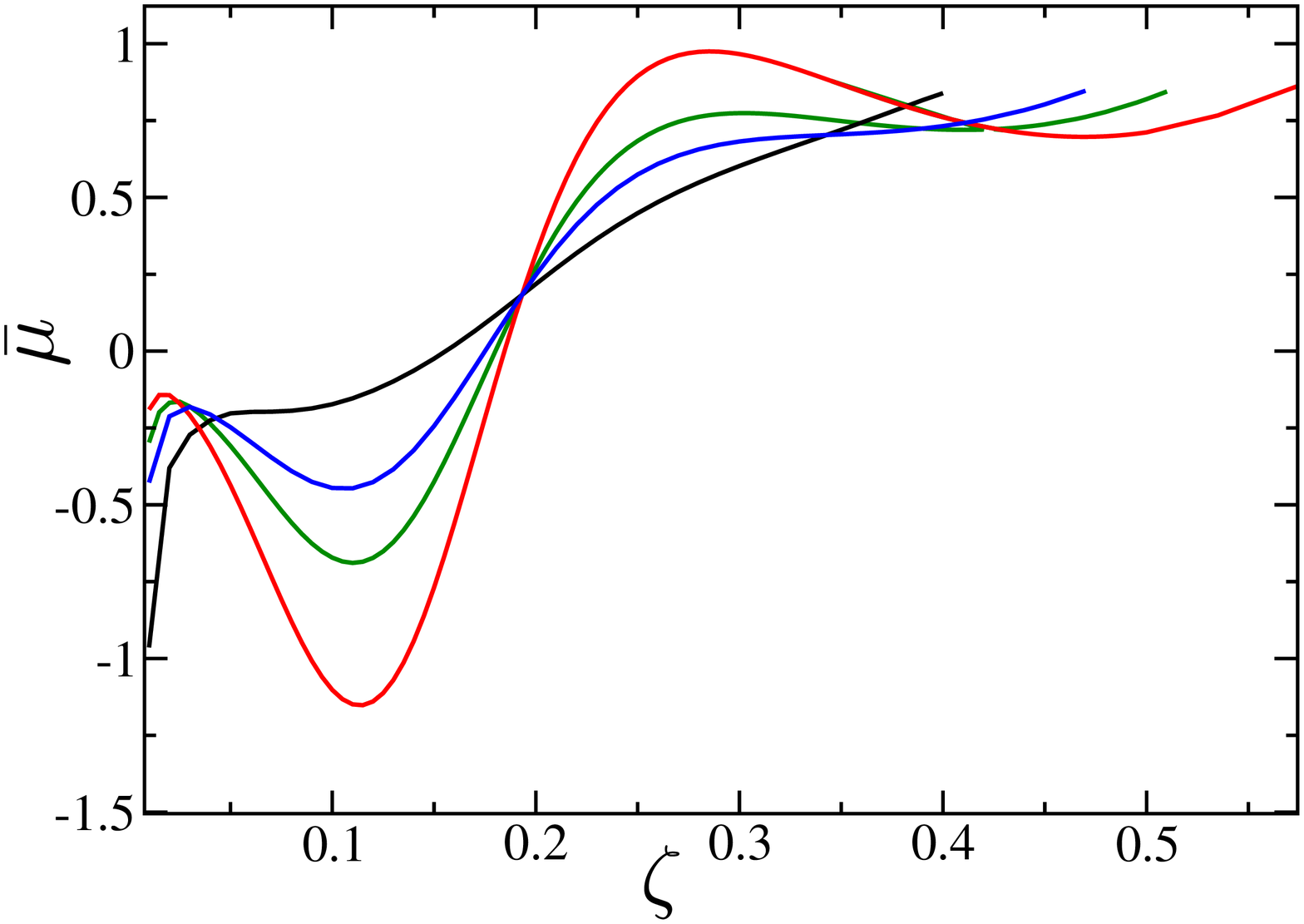}
 \includegraphics[scale=0.3]{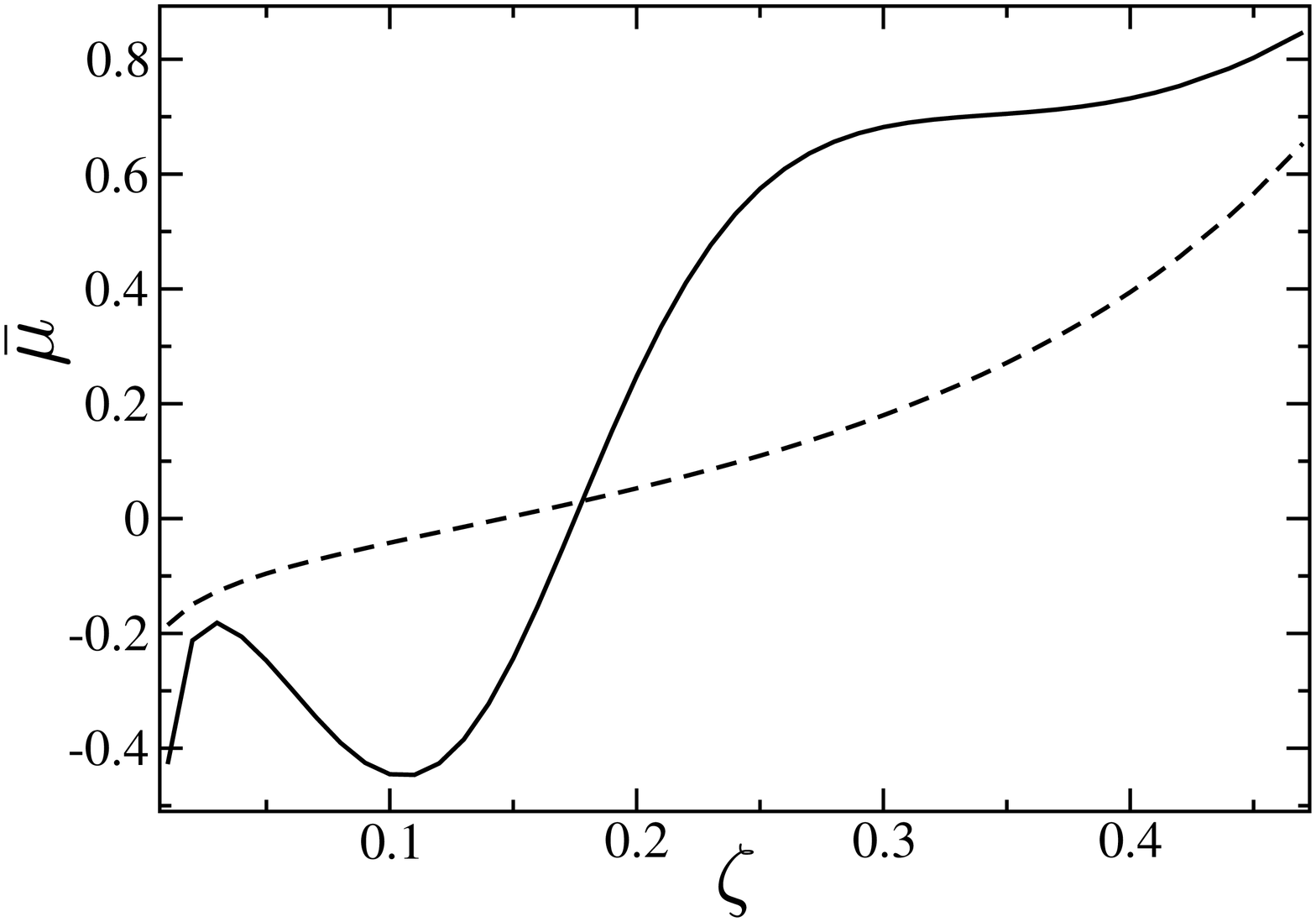}
 \caption{The chemical potential isotherms in $k_BT$ units for  System 2  as a function of the volume fraction (a) in our theory
 for $T^*=0.005, 0.007, 0.009, 0.015$ (solid lines, from bottom to top for $\zeta\approx 0.1$)
 and (b) in our theory (solid line) and in MF approximation, i.e. with the last two terms in Eq.(\ref{muhomma}) 
 neglected (dashed line) for $T^*=0.009$. }
 \label{muS2}
 \end{center}
\end{figure}

\begin{figure}
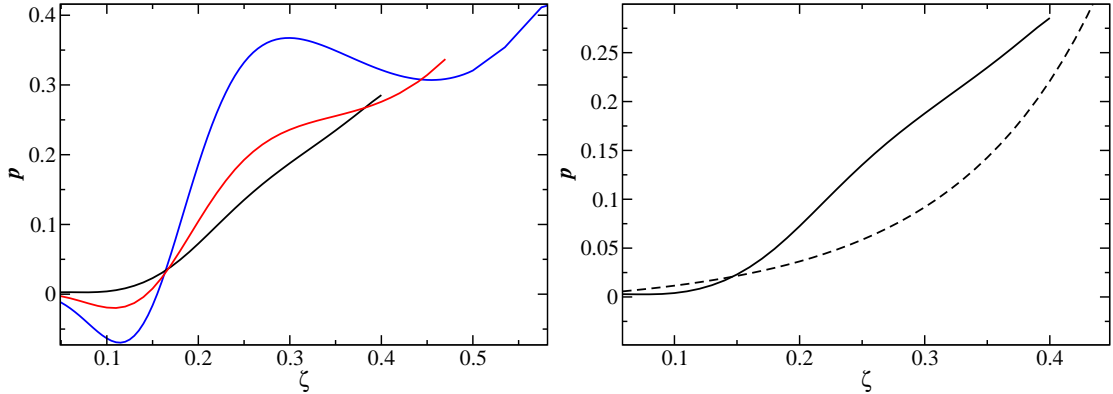

\begin{center}
 \includegraphics[scale=0.3]{fig15a.eps}
  \includegraphics[scale=0.3]{fig15b.eps}
 \caption{The pressure isotherms  in $k_BT/\sigma^3$ units as a function of the volume fraction  for  System 2 
 (a) in our theory, Eq.(\ref{pa})),
 for $T^*=0.005, 0.009, 0.015$ from top to bottom on the right and 
 (b) in our theory (solid line) and in MF (dashed line) for $T^*=0.015$. } 
 \label{pS2}
 \end{center}
\end{figure}

The shapes of the $\bar\mu(\zeta)$ and $p(\zeta)$ isotherms in System 2 are much more complex
than in System 1 (see Fig.\ref{muS2} and \ref{pS2}).
 For $T^*=0.015$ Eq.(\ref{criterion}) is satisfied for all $\zeta$. At this temperature we obtain very small slope of 
 $\bar\mu(\zeta)$ and $p(\zeta)$ (very large compressibility)
 for $\zeta\sim 0.05- 0.1$, and significantly larger  slope of  $\bar\mu(\zeta)$ and $p(\zeta)$ (smaller compressibility)
 that weakly depends on $\zeta$ for $\zeta>0.15$. When 
 $T^*$ is decreased to $0.009$,
 we obtain a van der Waals loop for $\zeta<0.15$, 
 very large  slope of  $\bar\mu(\zeta)$ and $p(\zeta)$ (very small compressibility) 
 for $\zeta\approx 0.2$ and very large compressibility for $\zeta>0.3$. 
 The theory becomes oversimplified for $\zeta<0.15$ for $T^*=0.009$ (see Fig.\ref{G0S2}), 
 and based on the comparison with the 1D model 
 we can expect that very large compressibility may be present  instead of the van der Wals loop obtained in our approximation.
 We can only conclude that 
 either very large compressibility or a phase transition occurs for
 low volume fractions for $T^*\le 0.009$.
 The very small compressibility for $\zeta\sim 0.2$ occurs together with the very large correlation length 
 of the correlation function that exhibits oscillatory decay.
 For decreasing $T^*$ the compressibility for $\zeta\sim 0.2$ significantly decreases, and the correlation length increases. 
  This behavior may indicate that  when $T^*$ decreases, the  clusters or layers of particles 
 become more and more ordered in space for  $\zeta\sim\zeta_c$, where the inhomogeneous 
 distribution of particles dominates. To compress such a system one has to decrease the separation between the aggregates,
 and overcome the repulsion between them.
 For $T^*\le 0.007$ the van der Waals loops are present
 for both small and large volume fractions. It is possible that phase transitions between the disordered gas- and liquid phases, 
 and the inhomogeneous phase stable for intermediate $\zeta$, occur for some range of $T^*$. 
 What remains unclear is the nature of the phase or phases with very small compressibility.
We considered only $\bar\zeta=const.$ and isotropic correlations, but  in reality
  periodically ordered phases, or phases with $\bar\zeta=const.$ and  anisotropic correlations, 
  may be stable for some temperature range for intermediate $\bar\zeta$.
  Further studies are necessary
 before drawing definite conclusions concerning the phase behavior in this 3D system. 
 Either pseudo phase transitions between the dilute and dense phases, and the inhomogeneous phase 
with periodic or quasi-periodic order,
or real phase transitions   can occur. In any case, when the van der Waals loops are obtained in this approximation, 
we can expect  a huge change of compressibility with increasing 
$\bar\zeta$.

\section{summary and discussion}
We have developed the theory for systems with inhomogeneities that form spontaneously on the mesoscopic length scale. 
In this work we have focused on the disordered phase, where the particles self-assemble into aggragates 
that do not form an ordered periodic pattern.
A hierarchy of equations relating the direct correlation functions $C_n$  with the correlation functions 
for the fluctuations of the local volume fraction of particles, $\phi({\bf r})$, has been constructed. 
To obtain a closed set of equations we have neglected the terms associated with $C_3$ in the equation 
for $\langle\phi({\bf r}_1)^2\phi({\bf r}_2)^2\rangle^{con}$. Next, neglecting the higher-order terms in the expression for $C_4$,
we have obtained and solved a self-consistent equation for $\langle\phi({\bf r}_1)^2\phi({\bf r}_2)^2\rangle^{con}$. 
This result leads finally to 
a closed set of equations for the two-point correlation and direct correlation functions, Eqs.(\ref{CG}), (\ref{selfconm0}), 
(\ref{Ds}) and (\ref{tildeD}).
Solutions of our self-consistent equations for the two-point functions allow us to 
calculate the fluctuation-corrections to the chemical potential $\bar\mu(\zeta)$  and pressure, Eqs.(\ref{muhomma}) and (\ref{pa}).

The general framework of our theory allows for systematic improvement of accuracy of the results. One can improve both, the approximation for 
the Boltzmann factor $\exp(-\beta H_f)$ that describes the probability of sponteneous appearence of $\phi({\bf r})$, 
and the approximation for $\langle\phi({\bf r}_1)^2\phi({\bf r}_2)^2\rangle^{con}$. In our theory $H_f$ is Taylor expanded in  $\phi$, 
and the expansion is truncated.
We have introduced a necessary condition for validity of a theory with  the Taylor
expansion truncated at the n-th order (see  Eq.(\ref{criterion}) for $n=4$).

We have applied our $\phi^4$ theory to the 1D lattice model with
competing interactions, and to two variants of the SALR model in 3D.
Following Ref.\cite{ciach:15:0}, we have postulated a functional form of the correlation function $G$ and have
solved numerically the equations for the parameters in the expression for $G$ that follow from our self-consistent equations
for the two-point correlation and direct correlation functions. Next, we
calculated  $\bar\mu(\zeta)$  and pressure for these models.

Comparison of the predictions of our approximate theory with the exact results obtained
for the 1D model in Ref.\cite{pekalski:13:0} allows for verification of the accuracy of our approximations.
The shape of the $\bar\mu(\zeta)$ and $p(\zeta)$ lines
agrees very well with exact results when the criterion (\ref{criterion}) is satisfied. 
All the qualitative trends are correctly reproduced. 
 However, we do not obtain quantitative agreement
at this level of approximation.
 When the criterion (\ref{criterion}) is not satisfied, i.e. for low $T^*$, van der Waals 
 loops in $\bar\mu(\zeta)$ are predicted in our theory,
  although in reality only pseudo phase transitions occur in this model.
 Thus, when the necessary condition (\ref{criterion}) is violated,  the theory is overesimplified indeed.
 We have made  preliminary calculations for $\bar\mu(\zeta)$ in the $\phi^6$ theory (not presented here)
and obtained very similar shapes of the lines, and reasonably good 
quantitative agreement with the exact results for $0.2<\zeta<0.8$. 
 In the $\phi^6$ theory
 the van der Waals loops appear for lower $T^*$ than in the $\phi^4$ theory. Thus, systematic improvement of
 accuracy of the results is indeed possible within the general framework of our theory. In order to obtain better accuracy for 
$ \zeta<0.2$ and $\zeta>0.8$, the term associated with $C_3$ should be included in Eq.(\ref{G4}).

  We applied our theory to two versions of the 3D SALR model (Eq.(\ref{int_pot_r}))
  described and discussed in sec.5. In System 1 the phase separation into homogeneous dilute and dense phases is strongly 
 unfavourable energetically and the periodic structure with the most probable
 period is only weakly favoured compared to the homogeneous structure. In
 System 2 the optimal periodic structure is strongly favourable, and the phase separation is weakly favourable energetically 
  compared to a homogeneous
  structure with given $\zeta$.
 The MF phase diagrams for the two considered versions of the 3D SALR model are nearly the same~\cite{ciach:10:1}.
 It is well known, however that the ordered phases are stable for much lower temperature than predicted by MF.
 Our aim was investigation of the structure and the EOS in the high-$T$ part of the MF instability region of the homogeneous phase,
 where  weakly ordered phases are predicted by MF, but in reality a disordered inhomogeneous phase is stable.
 
 Our results show the anomalous decrease of pressure for increasing $T$ for a range of $\zeta>\zeta_c$ in both systems. 
 Apart from this common feature,
 the effects of fluctuations depend very strongly on the shape of the interaction potential. In the repulsion-dominated System 1
 and in  System 2 where neither repulsion nor attraction dominates, the EOS is qualitatively
 different (see Figs.\ref{pS1} and \ref{pS2}).
 The disordering effects of fluctuations are much stronger in System 1. On the other hand, the shape of the
 $\mu(\zeta)$ and $p(\zeta)$ lines is much more complex in System 2.

 In System 1 we do not obtain any rapid change of the compressibility for increasing  $\zeta$ even for very small $T^*$.
 Unfortunately, for small $T^*$ the theory is overesimplified, especially for $\zeta<\zeta_c$,
 and we cannot draw definite conclusions concerning the phase transition between the dilute gas and inhomogeneous phases.
 For 
 $\zeta>\zeta_c$ we do not obtain large compressibility that would indicate an approach to a phase transition 
 between the  inhomogeneous phase and the dense liquid even for temperature as low as $T^*= 0.0015$ (in reduced units).

 In contrast, in System 2
 the compressibility changes rapidly for increasing $\zeta$ even for relatively high $T^*$, 
 and van der Waals loops  occur at sufficiently low $T^*$.
 The van der Waals loops suggest that phase transitions or pseudo phase transitions  between the gas and
 liquid phases, and the {\it disordered inhomogeneous phase} stable for intermediate volume fractions occur. 
 Based on the comparison with the 1D case, however, we expect that pseudo-phase transitions, indicating significant structural 
 changes in the disordered phase, occur when $T^*$ is high. 
 Note that  within the stability region of the disordered phase the gas, the  cluster fluid and the percolating fluid
  are distinguished in the recent simulations~\cite{zhuang:16:0}.
 We cannot rule out the possibility that for intermediate $T^*$ and $\zeta$ the  {\it disordered inhomogeneous phase}
  is a thermodynamically distinct phase that can coexist with the homogeneous dilute and dense phases.
  For low $T^*$ we may expect stability of the ordered periodic 
 phases for intermediate $\zeta$, and transitions between the disordered and the ordered phases.

 At present we can only speculate about the  phase behavior, because we have considered neither periodic
 phases nor phases with 
 orientational order (anisotropic correlation functions). 
 We cannot exclude phase transitions between inhomogeneous phases with different degree of order when $T^*$ decreases.
 However, such transitions cannot be investigated with the assumptions  of $\zeta=const.$ and isotropic correlations 
 that we have made in this work.
 We shall investigate phases with anisotropic correlations and the periodically ordered phases in our future studies.
 
 We conclude that the theory developed in this work can be a convenient tool for studying inhomogeneous systems. 
  Further studies
 are necessary for developing an approximation within our general framework 
 that would yield more accurate results on the quantitative level.
 
\acknowledgments
We   acknowledge the financial support by the NCN grant 2012/05/B/ST3/03302.

 \section{Appendices}
 \subsection{Explicit expressions for $C_3$ and $C_4$}
 
 From Eq.(\ref{Cn})  we obtain for $C_3$ and $C_4$
  \begin{eqnarray}
  \label{C3A}
    C_3({\bf r}_1,{\bf r}_2, {\bf r}_3)= A_3\delta({\bf r}_1,{\bf r}_2)\delta({\bf r}_2,{\bf r}_3)+\langle
    \frac{\delta^3 \beta H_f}{\delta\zeta(1)\delta\zeta(2)\delta\zeta(3)}
    \rangle
    \\\nonumber
   - \Bigg[\langle
   \frac{\delta^2 \beta H_f}{\delta\zeta(1)\delta\zeta(2)} \frac{\delta \beta H_f}{\delta\zeta(3)}
   \rangle^{con} + permut(2) \Bigg] +\langle
   \frac{\delta \beta H_f}{\delta\zeta(1)}\frac{\delta \beta H_f}{\delta\zeta(2)}\frac{\delta \beta H_f}{\delta\zeta(3)} \rangle^{con}
  \end{eqnarray}
and 

\begin{eqnarray}
\label{C4A}
    C_4({\bf r}_1,{\bf r}_2, {\bf r}_3, {\bf r}_4)= A_4\delta({\bf r}_1,{\bf r}_2)\delta({\bf r}_2,{\bf r}_3)\delta({\bf r}_3,{\bf r}_4)
    +\langle
    \frac{\delta^4 \beta H_f}{\delta\zeta(1)\delta\zeta(2)\delta\zeta(3)\delta\zeta(4)}
    \rangle
    \\\nonumber
   - \Bigg[\langle
   \frac{\delta^3 \beta H_f}{\delta\zeta(1)\delta\zeta(2)\delta\zeta(3)} \frac{\delta \beta H_f}{\delta\zeta(4)}
   \rangle^{con} + permut(3) \Bigg] 
    - \Bigg[\langle
   \frac{\delta^2 \beta H_f}{\delta\zeta(1)\delta\zeta(2)} \frac{\delta^2 \beta H_f}{\delta\zeta(3)\delta\zeta(4)}
   \rangle^{con} + permut(2) \Bigg]
    \\\nonumber
   +\Bigg[\langle
   \frac{\delta^2 \beta H_f}{\delta\zeta(1)\delta\zeta(2)}\frac{\delta \beta H_f}{\delta\zeta(3)}\frac{\delta \beta H_f}{\delta\zeta(4)} \rangle^{con} 
  + permut(5)   \Bigg]
   -\langle
   \frac{\delta \beta H_f}{\delta\zeta(1)}\frac{\delta \beta H_f}{\delta\zeta(2)}\frac{\delta \beta H_f}{\delta\zeta(3)}
   \frac{\delta \beta H_f}{\delta\zeta(4)} \rangle^{con}
  \end{eqnarray} 
  where we have simplified the notation introducing $\zeta(i)\equiv \zeta({\bf r}_i)$ and ``permut(n)'' means $n$ 
  different terms obtained by permutations of $(1,2,3)$ or $(1,2,3,4)$ in (\ref{C3A}) or (\ref{C4A}) respectively. 
$ \langle  X_1({\bf r}_1)... X_n({\bf r}_n) \rangle^{con}$ is the part of $ \langle  X_1({\bf r}_1)... X_n({\bf r}_n) \rangle$ 
that cannot be represented as a product  of average quantities calculated for disjoint sets of points.
When $H_f$ is
 Taylor-expanded, and only terms proportional to $A_n$ with $n\le 4$ are kept, we obtain the approximate expressions
 \begin{eqnarray}
  C_3({\bf r}',{\bf r}'', {\bf r}''')= A_3\delta({\bf r}',{\bf r}'')\delta({\bf r}'',{\bf r}''')
  \nonumber\\
  -\frac{A_3A_4}{4}\Bigg[\langle\phi({\bf r}')^2\phi({\bf r}'')^2\rangle^{con}\big(
  \delta({\bf r}',{\bf r}''')+\delta({\bf r}'',{\bf r}''')
  \big) 
  +\langle\phi({\bf r}')^2\phi({\bf r}''')^2\rangle^{con} \delta({\bf r}',{\bf r}'') \Bigg]
  \nonumber\\
   +
   \Big(\frac{A_3}{2}\Big)^3\langle\phi({\bf r}')^2\phi({\bf r}'')^2\phi({\bf r}''')^2\rangle^{con}
 \end{eqnarray}
and
\begin{eqnarray}
\label{C4appendix}
  C_4({\bf r}',{\bf r}'', {\bf r}''', {\bf r}'''')= 
  A_4\delta({\bf r}',{\bf r}'')\delta({\bf r}'',{\bf r}''')\delta({\bf r}''',{\bf r}'''')
  \nonumber\\
  - \Big(\frac{A_4}{2} \Big)^2\Bigg[\langle\phi({\bf r}')^2\phi({\bf r}'')^2\rangle^{con}\big(
  \delta({\bf r}',{\bf r}''')\delta({\bf r}'',{\bf r}'''')+\delta({\bf r}'',{\bf r}''')\delta({\bf r}',{\bf r}'''')
  \big)  \nonumber\\
  +\langle\phi({\bf r}')^2\phi({\bf r}''')^2\rangle^{con} \delta({\bf r}',{\bf r}'') \delta({\bf r}''',{\bf r}'''') \Bigg]
  \nonumber\\
   +
   \frac{A_4^2}{3!}\Bigg[\langle\phi({\bf r}')^3\phi({\bf r}'')\rangle^{con}\delta({\bf r}'',{\bf r}''') \delta({\bf r}'',{\bf r}'''') 
   + permut(3)\Bigg]
   \nonumber\\
   +
  \frac{A_4}{2} \Big(\frac{A_3}{2}\Big)^2
  \Bigg[\langle\phi({\bf r}')^2\phi({\bf r}'')^2\phi({\bf r}''')^2\rangle^{con}\delta({\bf r}',{\bf r}'''') +permut(2)\Bigg]
   \nonumber\\
   +\Big(\frac{A_3}{2}\Big)^4\langle\phi({\bf r}')^2\phi({\bf r}'')^2\phi({\bf r}''')^2\phi({\bf r}'''')^2\rangle^{con}
 \end{eqnarray}

\subsection{parameters $\xi_0,\alpha_0, A_0$ in Eq.(\ref{G3d}) and the expression for $\tilde D$ in 3D}

\begin{equation}
\label{A0}
 A_0^2=\frac{1}{(4\pi)^2v_0c_0},
\end{equation}
\begin{equation}
 \alpha_0^2 = \frac{c_0}{4v_0}\xi_0^2,
\end{equation}
\begin{equation}
\label{xi0}
 \xi_0^2= 2\frac{v_0k_0^2 +\sqrt{v_0^2k_0^4+v_0c_0}}{c_0},
\end{equation}
and for $G$  given in Eq.~(\ref{G3d})
$\tilde D(k)$ defined in Eq.~(\ref{tildeD}) takes the explicit form
\begin{equation}
 \tilde D(k)=\frac{\pi A_r^2}{k}
 \Bigg[
 2\arctan\Big( \frac{k\xi_r}{2}
 \Big) - \arctan\Big( \frac{(k+2\alpha_r)\xi_r}{2}\Big)  - \arctan\Big( \frac{(k-2\alpha_r)\xi_r}{2}\Big) 
 \Bigg].
\end{equation}

\end{document}